\RequirePackage{fix-cm}
\documentclass{svjour3}                     
\smartqed  

\usepackage{amsmath}
\usepackage{mathrsfs}
\usepackage{amssymb}            
\usepackage[english]{babel}
\usepackage[T1]{fontenc}
\usepackage[latin1]{inputenc}
\usepackage[dvips]{graphicx}
\usepackage{epsf,epsfig}
\usepackage[usenames]{color}
\usepackage{color,xcolor}
\usepackage{rotating}

\journalname{Celestial Mechanics and Dynamical Astronomy}
\begin{document}

\title{Phase space description of the dynamics due to the coupled effect of the planetary oblateness and the solar radiation pressure perturbations
}

\titlerunning{Oblateness and SRP induced dynamics}        

\author{Elisa Maria Alessi        \and
        Camilla Colombo \and
        Alessandro Rossi
}


\institute{E. M. Alessi, A. Rossi \at
             Istituto di Fisica Applicata ``Nello Carrara'' -- Consiglio Nazionale delle Ricerche (IFAC-CNR), Via Madonna del Piano 10, 50019 Sesto Fiorentino (FI), Italy \\
              Tel.: +39-0555226315\\
              \email{em.alessi@ifac.cnr.it, a.rossi@ifac.cnr.it}           
           \and
           C. Colombo \at
           Department of Aerospace Science and Technology, Politecnico di Milano, Via Giuseppe La Masa 34, 20156 Milano, Italy\\
             \email{camilla.colombo@polimi.it} 
}

\date{Received: date / Accepted: date}

\maketitle

\begin{abstract}
The aim of this work is to provide an analytical model to characterize the equilibrium points and the phase space associated with the singly-averaged dynamics caused by the planetary oblateness coupled with the solar radiation pressure perturbations. A two-dimensional differential system is derived by considering the classical theory, supported by the existence of an integral of motion comprising semi-major axis, eccentricity and inclination. Under the single resonance hypothesis, the analytical expressions for the equilibrium points in the eccentricity-resonant angle space are provided, together with the corresponding linear stability. The Hamiltonian formulation is also given. The model is applied considering, as example, the Earth as major oblate body, and a simple tool to visualize the structure of the phase space is presented. Finally, some considerations on the possible use and development of the proposed model are drawn.
\end{abstract}

\section{Introduction}
\label{sec:intro}

The main objective of this work is to derive an analytical model in the three-dimensional case of the equilibrium points associated with the singly-averaged dynamics induced by the coupled effect of the solar radiation pressure (SRP) and the planetary oblateness. As far as we know, such derivation does not exist at the moment and it can represent a fundamental tool in different fields. In particular,  the stationary configurations, the invariant curves in the libration regions corresponding to the elliptic points and the hyperbolic invariant manifolds associated with {\bf the} hyperbolic points can have several applications in planetary dynamics {\bf (e.g.,} \cite{Mi,REF9}) and in mission design {\bf (e.g.,} \cite{REF14,REF13}). 

The literature shows that the pioneers in the identification of the role of the SRP effect coupled with the zonal harmonics $J_2$ of a primary body in the orbital evolution of a small body were Musen \cite{pM60} and Cook \cite{gC62}. They developed the corresponding singly-averaged equations of motion, for a satellite orbiting the Earth, in terms of Keplerian orbital elements, and remarked the existence of six orbital resonances involving the rate of precession of the longitude of the ascending node, of the argument of pericenter and the apparent mean motion of the Sun. Later on, Hughes \cite{H77} expanded the disturbing function associated with SRP using the Kaula's method up to high order terms, and provided some examples on whether they can be relevant for satellites in Low Earth Orbit (LEO). Recall also that Breiter \cite{sB99,sB01} addressed the analytical treatment of luni-solar perturbations in canonical coordinates, considering not only the resonant condition, but more in general the critical points associated with the dynamical system, together with their stability\footnote{Recall that the SRP effect can be written analogously to the solar gravitational perturbation, except for the amplitude of the perturbation and the first order of its expansion \cite{H77}.}.

In terms of applications, in the vast literature on the dynamics of dust in planetary systems, we can mention Krivov et al. \cite{REF9} and Hamilton and Krivov \cite{REF10}, who studied the eccentricity oscillation associated with the singly-averaged disturbing potential of SRP and $J_2$ for a dust particle orbiting a planet. Krivov and Getino \cite{REF11} applied the same model, but in the planar case and only considering one of the resonant terms, {\bf also} for obtaining a phase space representation of the orbit evolution of low-inclination Earth orbits, in the perspective of high-altitude balloons. 

\noindent Colombo et al. \cite{REF1} performed a parametric analysis of the SRP-$J_2$ phase space for different values of semi-major axis and area-to-mass ratio, identifying the equilibrium points corresponding to either heliotropic (apogee pointing towards the Sun) or anti-heliotropic (perigee pointing towards the Sun) orbits. The equilibrium conditions computed analytically for the planar case were numerically extended to non-zero inclinations. For such frozen orbits, different applications were proposed: the use of heliotropic orbits for Earth observation in the visible wavelength \cite{REF1,REF2}, or the use of heliotropic orbits at an oblate asteroid \cite{REF12}, or also the use of anti-heliotropic orbits for geomagnetic tail exploration missions \cite{REF7,REF6,REF8}. Following  \cite{REF11} a particular application was proposed by L\"ucking at al. \cite{REF4,REF5} for passive deorbiting of spacecraft at the end-of-life. Given the mass parameter of the main body, the dynamics under consideration depends on three constants, the semi-major axis of the orbit, a given integral of motion and the area-to-mass ratio of the small body. In \cite{REF4,REF5,REF15} the area-to-mass required to obtain a given eccentricity increase was computed by means of an optimization procedure, considering also the associated time. 

\noindent With the same aim of looking for natural perturbations that can support a reentry from LEO at the end-of-life, Alessi et al. \cite{Alessi_CMDA} performed an extensive numerical mapping of the region, and identified deorbiting natural corridors, defined in terms of inclination and eccentricity for given semi-major axis. In \cite{Schettino_freq}, a numerical frequency portrait of the LEO region was presented, highlighting also the role that high-order resonances associated with SRP \cite{H77} might have. 

\noindent {\bf The other typical example of mission that can be designed on the basis of the coupled SRP-$J_2$ dynamics is a mission to an asteroid or a comet (e.g, \cite{Russell,Scheeres}).}

In this work, the three-dimensional equations of motion which describe the variation in eccentricity, inclination, and a given angle accounting for the motion of the longitude of the ascending node and the argument of pericenter are analyzed, they are reduced to a two-dimensional case by means of a well-known integral of motion and the analytical expressions for resonant conditions and equilibrium points are provided. The corresponding stability is given for the six main resonances and the possible different phase space portraits for orbits at the Earth are described. Finally, the tools presented are analyzed in the perspective of a practical exploitation. 

Throughout the work, it is always assumed that the major body is the Earth, but the analytical treatment developed is general and it can be applied to any oblate body. {\bf In our formulation, we adopted the Earth equatorial plane as reference plane, because our major interest is a possible exploitation for deorbiting. We notice, however, that for other applications, e.g., \cite{Scheeres}, the ecliptic reference system can be more suitable. 

Moreover, the concept proposed will hold certainly in the cases described above, where the two perturbations -- oblateness and solar radiation pressure -- play a major role, but it will be of high interest to see how the solutions presented may persist under the effect of an additional effect and under which conditions they remain the skeleton of the long-term dynamics. For orbits at the Earth, the key case will be the dynamics due to lunisolar gravitational perturbations for Highly Elliptical Orbits \cite{Camilla2019} and in the geosynchronous region (e.g., \cite{CasanovaEtAl,GkoliasColombo}).}

\section{Dynamical Model}
\label{sec:model}

Let us assume that a small body (e.g., a spacecraft) moves under the effect of the Earth's gravitational monopole, the Earth's oblateness and the solar radiation pressure. In particular, for the SRP it is assumed the cannonball model, that the orbit of the spacecraft is entirely in sunlight and that the effect of the Earth's albedo is negligible (e.g., \cite{REF9}). 

In the following, $(a,e,i,\Omega,\omega)$ are the Keplerian orbital elements of the small body measured with respect to the Earth equatorial plane, $n$ the mean motion of the small body, $\mu$ the Earth gravitational parameter, $J_2$ the second zonal term of the geopotential, $r_{\oplus}$  the equatorial radius of the
Earth, $C_{\oplus}=\mu J_2r^2_{\oplus}$, $\epsilon$ the obliquity of the ecliptic, $P$ the solar radiation pressure at 1 AU, $c_R$ the reflectivity coefficient\footnote{We always assume $c_R=1$ in this work.}, $A/m$ the area-to-mass ratio of the small body, $C_{SRP}=\frac{3}{2}Pc_R\frac{A}{m}$, $\lambda_S$ the longitude of the Sun measured on the ecliptic plane, and {\bf the resonant argument is}
\begin{equation}\label{eq:psi}
\psi_j=n_1\Omega+n_2\omega + n_3 \lambda_S, \quad \quad \quad \psi_j\in[0,2\pi],
\end{equation}
with $n_1, n_2, n_3$ are as in Table~\ref{tab:dotpsij}, where we provide also the correspondence with respect to the notation used by Cook \cite{gC62}.

Under these hypotheses, following the description given in \cite{REF9,REF1,REF2,Alessi_MNRAS}, 
the singly-averaged equations of motion of the spacecraft can be written as
\begin{equation}\label{eq:equmot}
\begin{aligned}
\frac{da}{dt}&=0,\\
\frac{de}{dt}&=-C_{SRP}\frac{\sqrt{1-e^2}}{na}\sum_{j=1}^6\mathcal{T}_{j}{\partial{\cos{\psi}_j}\over\partial\omega},\\
\frac{di}{dt}&=-C_{SRP}\frac{e}{na\sqrt{1-e^2}\sin i}\sum_{j=1}^6\mathcal{T}_{j}\left({\partial{\cos{\psi}_j}\over\partial \Omega}-\cos i{\partial{\cos{\psi}_j}\over\partial \omega}\right),\\
\frac{d\Omega}{dt}&=\dot\Omega_{J_2}+\dot\Omega_{SRP},\\
\frac{d\omega}{dt}&=\dot\omega_{J_2}+\dot\omega_{SRP},
\end{aligned}
\end{equation}
where
\begin{equation}\label{eq:Tj}
\begin{aligned}
\mathcal{T}_1&=\cos ^2\left(\frac{\epsilon}{2}\right) \cos ^2\left(\frac{i}{2}\right), \\
\mathcal{T}_2&=\cos^2 \left(\frac{\epsilon}{2}\right) \sin ^2\left(\frac{i}{2}\right),\\
\mathcal{T}_3&=\frac{1}{2} \sin (\epsilon) \sin (i), \\
\mathcal{T}_4&=-\frac{1}{2} \sin (\epsilon) \sin (i),\\
\mathcal{T}_5&=\sin ^2\left(\frac{\epsilon}{2}\right) \cos ^2\left(\frac{i}{2}\right),\\
\mathcal{T}_6&=\sin ^2\left(\frac{\epsilon}{2}\right) \sin ^2\left(\frac{i}{2}\right),
\end{aligned}
\end{equation}
and
\begin{equation}\label{bompom}
\begin{aligned}
\dot\Omega_{J_2} &= -\frac{3}{2}\frac{J_2r^2_{\oplus}n}{a^2(1-e^2)^2}\cos i,\\
\dot\Omega_{SRP}&=C_{SRP}\frac{e}{na\sqrt{1-e^2}\sin i}\sum_{j=1}^6{\partial{\mathcal{T}}_{j}\over\partial i}\cos{\psi}_j,\\
\dot\omega_{J_2} &= \frac{3}{4}\frac{J_2r^2_{\oplus}n}{a^2(1-e^2)^2}\left(5\cos^2i-1\right),\\
\dot\omega_{SRP}&=C_{SRP}\frac{\sqrt{1-e^2}}{nae}\sum_{j=1}^6\mathcal{T}_{j}\cos{\psi}_j-\dot\Omega_{SRP}\cos i.
\end{aligned}
\end{equation}

\begin{table}
\caption{Argument  $\psi_j=n_1\Omega+n_2\omega + n_3 \lambda_S$ of the periodic component in terms of $n_1, n_2, n_3$. In the table, it is showed also the corresponding notation in the work by Cook \cite{gC62}.\vspace{5mm}}
\label{tab:dotpsij}
\begin{tabular}{cccc|c} 
\hline\noalign{\smallskip}
$j$ & $n_1$ & $n_2$ & $n_3$ & Cook\\
\noalign{\smallskip}\hline\noalign{\smallskip}
		1 & 1 & ~1 & -1 & 8\\
		2 & 1 & -1 & -1& 7 \\
		3 & 0 & ~1 & -1& 15\\
		4 & 0 & ~1 & ~1& 14\\
		5 & 1 & ~1 & ~1& 6\\
		6 & 1 & -1 & ~1& 9\\
\noalign{\smallskip}\hline
\end{tabular}
\end{table}

If we assume that the dynamics is driven by a single term $j$ at a time, that means that only one periodic component $\sin\psi_j$ (for $e,i$; $\cos\psi_j$ for $\Omega, \omega$) affects the motion at a time, then we can simplify the description of the dynamics by analyzing the following system 
\begin{equation}\label{eq:red}
\begin{aligned}
\left. \frac{de}{dt} \right\rvert_{j}&=n_2C_{SRP}\frac{\sqrt{1-e^2}}{na}\mathcal{T}_{j}\sin{\psi}_j,\\
\frac{d\psi_j}{dt}&=n_1\dot\Omega_{(J_2,j)}+n_2\dot\omega_{(J_2,j)}+n_3n_S,
\end{aligned}
\end{equation}
where $n_S$ is the apparent mean motion of the Sun around the Earth, and $\dot\Omega_{(J_2,j)}$ and $\dot\omega_{(J_2,j)}$ are intended to include the contribution of $J_2$ and the $j$th-contribution of the SRP. In the following, the time derivative of the eccentricity due to the periodic term $j$, namely, $\left. \frac{de}{dt} \right\rvert_{j}$, will appear as $\dot e_{|j}$.

\noindent 

{\bf Since we can apply the same argument used in \cite{D16} for the coupled lunisolar gravitational perturbation and oblateness effect (see \cite{Alessi_MNRAS} and Sec. \ref{sec:ham}), }
the behavior in inclination can be recovered at any time by means of the integral of motion
\begin{equation}\label{eq:int}
\Lambda=(n_2\cos{i}-n_1)\sqrt{\mu a(1-e^2)},
\end{equation}
which can be inverted to get
\begin{equation}\label{eq:cosi}
\cos{i}=\frac{\Lambda}{n_2\sqrt{\mu a(1-e^2)}}+\frac{n_1}{n_2}.
\end{equation}
Note that the evolution of the dynamics is determined by three parameters, the semi-major axis $a$, the area-to-mass ratio $A/m$ and the integral of motion $\Lambda$ (which depends on the dominant $j-$resonant term). 

\subsection{Resonances and Equilibrium Points}

A resonance takes place when the following condition is satisfied 
\begin{equation}\label{eq:res_cond}
\dot\psi_j= 0.
\end{equation}
{\bf Notice that this condition implies specific configurations, according to the periodic component $j$ considered. In particular, 
\begin{itemize}
\item for $j=1$, if Eq.~(\ref{eq:res_cond}) is verified at prograde orbits, then the longitude of the periapsis is sun-synchronous; 
\item for $j=2$, if Eq.~(\ref{eq:res_cond}) is verified at retrograde orbits, then the longitude of the periapsis is sun-synchronous;
\item for $j=3$, the argument of periapsis is sun-synchronous;
\item for $j=4$, the argument of periapsis is sun-antisynchronous;
\item for $j=5$, if Eq.~(\ref{eq:res_cond}) is verified at prograde orbits, then the longitude of the periapsis is sun-antisynchronous; 
\item for $j=6$, if Eq.~(\ref{eq:res_cond}) is verified at retrograde orbits, then the longitude of the periapsis is sun-antisynchronous. 
\end{itemize}}

Note that Eq.~(\ref{eq:res_cond}) depends on $(a,e,i,\psi_j)$, that is, also on $C_{SRP}$, except if it is computed at $\psi_j=\pi/2$ or $\psi_j=3\pi/2$. In the latter case, the resonant condition can be computed only considering the effect due to $J_2$ on $(\Omega, \omega)$, that is, given $(a,e,n_1,n_2,n_3)$, by solving the following quadratic equation in $\cos i$
\begin{equation}\label{eq:res}
c_1\cos^2{i}+c_2\cos{i}+c_3=0,
\end{equation}
where
\begin{equation}\label{eq:coef_res}
\begin{aligned}
c_1&=\frac{15n_2J_2r^2_{\oplus}n}{4a^2(1-e^2)^2},\\
c_2&=-\frac{3n_1J_2r^2_{\oplus}n}{2a^2(1-e^2)^2},\\
c_3&=n_3n_S-\frac{3n_2J_2r^2_{\oplus}n}{4a^2(1-e^2)^2}.
\end{aligned}
\end{equation}
These are the resonant conditions depicted by Cook \cite{gC62} in his Fig.~5 and remarked in \cite{Alessi_CMDA,Schettino_freq,Alessi_MNRAS}. 
 
On the other hand, an equilibrium point is computed if both equations in Eqs.~(\ref{eq:red}) cancel out. This must happen in particular at $\psi_j=0$ or $\psi_j=\pi$. For $j=1,2,5,6$, discarding the singularity at $e=0$ and $e=1$, the equation that gives the value of inclination corresponding to the location of the equilibrium points at $\psi_j=0$  and at $\psi_j=\pi$, given $(a,e,A/m)$, is also a quadratic equation -- Eq.~(\ref{eq:res}) -- in $\cos{i}$ that can be solved. The corresponding analytical coefficients are showed in Tabs.~\ref{tab:quadeqeq0}-\ref{tab:quadeqeqpi} for all the cases. For $j=3,4$, discarding also the singularity corresponding to $i=0$, the equation is instead a cubic equation of the type $\sin^3{i}+s_1\sin^2{i}+s_2\sin{i}+s_3=0$, that can also be solved analytically via classical techniques. In Tabs.~\ref{tab:cubeqeq0}-\ref{tab:cubeqeqpi}, we show the coefficients of such cubic equations.

\begin{table}
\caption{Coefficients of the quadratic equation $c_1\cos^2{i}+c_2\cos{i}+c_3=0$ associated with the equilibrium point at $\psi_j=0$ for $j=1,2,5,6$. In the table, $C_{\oplus}=\mu J_2r^2_{\oplus}$, $C_{SRP}=\frac{3}{2}Pc_R\frac{A}{m}$, $\beta=\sqrt{1-e^2}$, $\gamma=\cos ^2\left(\frac{\epsilon}{2}\right)$, $\rho=\sin ^2\left(\frac{\epsilon}{2}\right)$. \vspace{5mm}}
\label{tab:quadeqeq0}
\begin{tabular}{llll} 
\hline\noalign{\smallskip}
$j$ & $c_1$ & $c_2$ & $c_3$\\
\noalign{\smallskip}\hline\noalign{\smallskip}
		1 & $15C_{\oplus}e\beta$ & $-6C_{\oplus}e\beta+2C_{SRP}a^4\beta^4\gamma$ & $~~2C_{SRP}a^4\beta^4\gamma(1-2e^2)-e\beta\left(3C_{\oplus}+4a^5\beta^4nn_S\right)$\\
		2 & $-15C_{\oplus}e\beta$  & $-6C_{\oplus}e\beta+2C_{SRP}a^4\beta^4\gamma$ & $-2C_{SRP}a^4\beta^4\gamma(1-2e^2)+e\beta\left(3C_{\oplus}-4a^5\beta^4nn_S\right)$ \\
		5 &  $15C_{\oplus}e\beta$ & $-6C_{\oplus}e\beta+2C_{SRP}a^4\beta^4\rho$  & $~~2C_{SRP}a^4\beta^4\rho(1-2e^2)-e\beta\left(3C_{\oplus}-4a^5\beta^4nn_S\right)$\\
		6 &  $-15C_{\oplus}e\beta$ & $-6C_{\oplus}e\beta+2C_{SRP}a^4\beta^4\rho$ & $-2C_{SRP}a^4\beta^4\rho(1-2e^2)+e\beta\left(3C_{\oplus}+4a^5\beta^4nn_S\right)$ \\
\noalign{\smallskip}\hline
\end{tabular}
\end{table}

\begin{table}
\caption{Coefficients of the quadratic equation $c_1\cos^2{i}+c_2\cos{i}+c_3=0$ associated with the equilibrium point at $\psi_j=\pi$ for $j=1,2,5,6$. In the table, $C_{\oplus}=\mu J_2r^2_{\oplus}$, $C_{SRP}=\frac{3}{2}Pc_R\frac{A}{m}$, $\beta=\sqrt{1-e^2}$, $\gamma=\cos ^2\left(\frac{\epsilon}{2}\right)$, $\rho=\sin ^2\left(\frac{\epsilon}{2}\right)$.\vspace{5mm}}
\label{tab:quadeqeqpi}
\begin{tabular}{llll} 
\hline\noalign{\smallskip}
$j$ & $c_1$ & $c_2$ & $c_3$\\
\noalign{\smallskip}\hline\noalign{\smallskip}
		1 & $15C_{\oplus}e\beta$ & $-6C_{\oplus}e\beta-2C_{SRP}a^4\beta^4\gamma$ & $-2C_{SRP}a^4\beta^4\gamma(1-2e^2)-e\beta\left(3C_{\oplus}+4a^5\beta^4nn_S\right)$\\
		2 &   $-15C_{\oplus}e\beta$  & $-6C_{\oplus}e\beta-2C_{SRP}a^4\beta^4\gamma$ & $~~2C_{SRP}a^4\beta^4\gamma(1-2e^2)+e\beta\left(3C_{\oplus}-4a^5\beta^4nn_S\right)$ \\
		5 &  $15C_{\oplus}e\beta$ & $-6C_{\oplus}e\beta-2C_{SRP}a^4\beta^4\rho$ & $-2C_{SRP}a^4\beta^4\rho(1-2e^2)-e\beta\left(3C_{\oplus}-4a^5\beta^4nn_S\right)$ \\
		6 &   $-15C_{\oplus}e\beta$ & $-6C_{\oplus}e\beta-2C_{SRP}a^4\beta^4\rho$  & $~~2C_{SRP}a^4\beta^4\rho(1-2e^2)+e\beta\left(3C_{\oplus}+4a^5\beta^4nn_S\right)$ \\
\noalign{\smallskip}\hline
\end{tabular}
\end{table}

\begin{table}
\caption{Coefficients of the cubic equation $\sin^3{i}+s_1\sin^2{i}+s_2\sin{i}+s_3=0$ associated with the equilibrium point at $\psi_j=0$ for $j=3$ and $j=4$. In the table, $C_{\oplus}=\mu J_2r^2_{\oplus}$, $C_{SRP}=\frac{3}{2}Pc_R\frac{A}{m}$, $\beta=\sqrt{1-e^2}$. \vspace{5mm}}
\label{tab:cubeqeq0}
\begin{tabular}{llll} 
\hline\noalign{\smallskip}
$j$ & $s_1$ & $s_2$ & $s_3$\\
\noalign{\smallskip}\hline\noalign{\smallskip}
		3 & $-\frac{2C_{SRP}}{15C_{\oplus}e}a^4\beta^3\sin{\epsilon}$ & $-\frac{4}{15C_{\oplus}}\left(3C_{\oplus}-a^5\beta^4n n_s\right)$ & $\frac{2C_{SRP}}{15C_{\oplus}}ea^4\beta^3\sin{\epsilon}$\\
		4 &  $ \frac{2C_{SRP}}{15C_{\oplus}e}a^4\beta^3\sin{\epsilon}$  & $-\frac{4}{15C_{\oplus}}\left(3C_{\oplus}+a^5\beta^4n n_s\right)$ & $-\frac{2C_{SRP}}{15C_{\oplus}}ea^4\beta^3\sin{\epsilon}$ \\
\noalign{\smallskip}\hline
\end{tabular}
\end{table}

\begin{table}
\caption{Coefficients of the cubic equation $\sin^3{i}+s_1\sin^2{i}+s_2\sin{i}+s_3=0$ associated with the equilibrium point at $\psi_j=\pi$ for $j=3$ and $j=4$. In the table, $C_{\oplus}=\mu J_2r^2_{\oplus}$, $C_{SRP}=\frac{3}{2}Pc_R\frac{A}{m}$, $\beta=\sqrt{1-e^2}$. \vspace{5mm}}
\label{tab:cubeqeqpi}
\begin{tabular}{llll} 
\hline\noalign{\smallskip}
$j$ & $s_1$ & $s_2$ & $s_3$\\
\noalign{\smallskip}\hline\noalign{\smallskip}
		3 & $\frac{2C_{SRP}}{15C_{\oplus}e}a^4\beta^3\sin{\epsilon}$ & $-\frac{4}{15C_{\oplus}}\left(3C_{\oplus}-a^5\beta^4n n_s\right)$ & $-\frac{2C_{SRP}}{15C_{\oplus}}ea^4\beta^3\sin{\epsilon}$\\
		4 &  $ -\frac{2C_{SRP}}{15C_{\oplus}e}a^4\beta^3\sin{\epsilon}$  & $-\frac{4}{15C_{\oplus}}\left(3C_{\oplus}+a^5\beta^4n n_s\right)$ & $\frac{2C_{SRP}}{15C_{\oplus}}ea^4\beta^3\sin{\epsilon}$ \\
\noalign{\smallskip}\hline
\end{tabular}
\end{table}

The stability of the points can be evaluated by computing the eigenvalues of the matrix
\begin{equation}\label{eq:jac}
J=\begin{pmatrix}
{\partial{\dot e|j}\over\partial e} & {\partial{\dot e|j}\over\partial{\psi_j}}  \\
{\partial{\dot \psi_j}\over\partial e} & {\partial{\dot \psi_j} \over\partial{ \psi_j}}
\end{pmatrix},
\end{equation}
at the given equilibrium point, where 
\begin{equation}\label{eq:partial}
\begin{aligned}
{\partial{\dot e_{|j}}\over\partial e} &=0,\\
{\partial{\dot e_{|j}}\over\partial{\psi_j}}  &=n_2C_{SRP}\frac{\sqrt{1-e^2}}{na}\mathcal{T}_{j}\cos{\psi}_j,\\
{\partial{\dot \psi_j}\over\partial e} &=n_1{\partial{\dot \Omega_{(J_2,j)}}\over\partial e}+n_2{\partial{\dot \omega_{(J_2,j)}}\over\partial e},\\
 {\partial{\dot \psi_j} \over\partial{ \psi_j}}&=0.
 \end{aligned}
\end{equation}
Note that the partial derivatives appearing in the third equation of Eqs.~(\ref{eq:partial}) can be evaluated by taking into account Eq.~(\ref{eq:cosi}), namely, 
{\footnotesize
\begin{equation*}
\begin{aligned}
{\partial{\dot \Omega_{(J_2,j)}}\over\partial e}&=-6\frac{J_2r^2_{\oplus}ne}{a^2(1-e^2)^3}\cos i+\frac{C_{SRP}}{na(1-e^2)^{3/2}\sin i}{\partial{\mathcal{T}}_{j}\over\partial i}\cos{\psi}_j-\frac{3}{2}\frac{J_2r^2_{\oplus}n}{a^2(1-e^2)^2}{\partial{\cos{i}}\over\partial e}+\\
&C_{SRP}\frac{e}{na\sqrt{1-e^2}}{\partial{\mathcal{T}}_{j}\over\partial i}\cos{\psi}_j{\partial\over\partial e}\frac{1}{\sin{i}}+C_{SRP}\frac{e}{na\sqrt{1-e^2}\sin i}\cos{\psi}_j{\partial\over{\partial e}}{{\partial{\mathcal{T}}_{j}\over\partial i}},\\
{\partial{\dot \omega_{(J_2,j)}}\over\partial e}&=3\frac{J_2r^2_{\oplus}ne}{a^2(1-e^2)^3}(5\cos^2{i}-1)-\frac{C_{SRP}}{na\sqrt{1-e^2}}\left(\frac{\mathcal{T}_{j}}{e^2}-\frac{\cos{i}}{(1-e^2)^2\sin i}{\partial{\mathcal{T}}_{j}\over\partial i}\right)\cos{\psi}_j+\\
& \frac{15}{2}\frac{J_2r^2_{\oplus}n\cos{i}}{a^2(1-e^2)^2}{\partial{\cos{i}}\over\partial e}+C_{SRP}\frac{\sqrt{1-e^2}}{nae}\cos{\psi}_j{\partial{\mathcal{T}_{j}}\over\partial{e}}-\\
&C_{SRP}\frac{e}{na\sqrt{1-e^2}}{\partial{\mathcal{T}}_{j}\over\partial i}\cos{\psi}_j{\partial\over\partial e}\frac{1}{\tan{i}}+C_{SRP}\frac{e}{na\sqrt{1-e^2}\tan{i}}\cos{\psi}_j{\partial\over{\partial e}}{{\partial{\mathcal{T}}_{j}\over\partial i}},
 \end{aligned}
\end{equation*}
}
where
\begin{eqnarray*}
{\partial{\cos{i}}\over\partial e}&=&\frac{e\Lambda}{n_2\sqrt{\mu a}(1-e^2)^{3/2}},\\
{\partial{\sin{i}}\over\partial e}&=&\frac{1}{\tan{i}}{\partial{\cos{i}}\over\partial e},
\end{eqnarray*}
and ${\partial{\mathcal{T}}_{j}\over\partial i}$ and ${\partial\over{\partial e}}{{\partial{\mathcal{T}}_{j}\over\partial i}}$ can be derived from these expressions by applying the half-angle trigonometric formulae to Eqs.~(\ref{eq:Tj}).

\subsection{Hamiltonian formulation}\label{sec:ham}

For completeness, we provide also the Hamiltonian formulation of the dynamics under study. To this end, {\bf we first recall that the equations of motion Eqs.~(\ref{eq:equmot}) and~(\ref{bompom}) are obtained from the singly-averaged disturbing potential associated with the oblateness and the solar radiation pressure effect, as described in  \cite{REF9,REF1,REF2,Alessi_MNRAS}.  Notice that the disturbing potential associated with the SRP also corresponds to Eq.~(17) in \cite{CasanovaEtAl} and to Eq.~(69) in \cite{Scheeres}. Then,} we can apply the argument developed in \cite{D16} for lunisolar gravitational perturbations, adapted to the solar radiation pressure effect, under the assumption that only one periodic term $j$ is relevant at a time for the motion of the small body. Considering, in particular, one of the two canonical transformations developed in \cite{D16} written in terms of classical Delaunay variables $\left(L, G, H, l, g, h \right)$, {\bf by introducing the canonical variables $\left(\Gamma, \tau\right)$ such that $\dot\tau\equiv {\partial{\mathcal{H}}\over\partial\Gamma}=\lambda_S$}, we can formulate the Hamiltonian of the problem in action angle variables $(\Sigma, \sigma)$. We have\footnote{In our formulation, $n_1$ and $n_2$ are inverted with respect to the formulation in \cite{D16} and we call $\Sigma_{1,2,3}$ what they call $\Lambda_{1,2,3}$.}
\begin{equation}\label{eq:lambdasigma}
\begin{aligned}
\Sigma_1&\equiv\Sigma=n_2^{-1}G, \quad \quad \quad   \quad \quad~ \sigma_1\equiv\sigma=n_2g+n_1h+n_3\tau,\\
\Sigma_2&\equiv\Lambda=-n_1G+n_2H, \quad \quad ~\sigma_2=n_2^{-1}h, \\
\Sigma_3&= -n_2^{-1}n_3G+\Gamma, \quad \quad \quad~~ \sigma_3=\tau.
 \end{aligned}
\end{equation}
Note that $\sigma_1\equiv\sigma\equiv\psi_j$ and that Eq.~(\ref{eq:int}) is derived from $\Sigma_2$.

In this way, we obtain
\begin{equation}\label{eq:ham}
\mathcal{H}=\mathcal{H}_{J_2}+\mathcal{H}_{SRP},
\end{equation}
where the part associated with $J_2$ is
\begin{equation}\label{eq:ham_j2}
\mathcal{H}_{J_2}=\frac{J_2 \mu^4 r^2_{\oplus} \left(-3 \Sigma_1^2 n_1^2 n_2^2+\Sigma_1^2 n_2^4-6 \Sigma_1 \Sigma_2 n_1 n_2-3 \Sigma_2^2\right)}{4 L^3 \Sigma_1^5 n_2^7},
\end{equation}
and the one due to SRP is
\begin{equation}\label{eq:ham_SRP}
\mathcal{H}_{SRP}=-\frac{C_{SRP}}{2\mu} L^2  \mathcal{S}_j\sqrt{1-\frac{n_2^2\Sigma_1^2}{L^2}} \cos\sigma_1+n_S\left(n_3\Sigma_1+\Sigma_3\right),
\end{equation}
where $\mathcal{S}_j$ depends on the resonance, namely,
\begin{equation}\label{eq:Sj}
\begin{aligned}
\mathcal{S}_1&=\cos^2\left(\frac{\epsilon}{2}\right)\left(\text{ci}+1\right),\\
\mathcal{S}_2&=\cos^2\left(\frac{\epsilon}{2}\right)\left(-\text{ci}+1\right),\\
\mathcal{S}_3&=\sin\epsilon\sqrt{-\text{ci}^2+1},\\
\mathcal{S}_4&=-\sin\epsilon\sqrt{-\text{ci}^2+1},\\
\mathcal{S}_5&=\sin^2\left(\frac{\epsilon}{2}\right)\left(\text{ci}+1\right),\\
\mathcal{S}_6&=\sin^2\left(\frac{\epsilon}{2}\right)\left(-\text{ci}+1\right),
 \end{aligned}
\end{equation}
being  $\text{ci}=\frac{n_1n_2\Sigma_1+\Sigma_2}{n_2^2\Sigma_1}$.

This is also to say that the equations of motion can be also written as
\begin{equation}\label{eq:eqmot_can}
\begin{aligned}
\dot\Sigma&=-{\partial{\mathcal{H}\left(\Sigma,\sigma\right)}\over\partial{\sigma}},\\
\dot\sigma&={\partial{\mathcal{H}\left(\Sigma,\sigma\right)}\over\partial{\Sigma}}.
 \end{aligned}
\end{equation}
Note that the Hamiltonian function is a constant of motion and so it is $\Sigma_3$, since $\mathcal{H}$ does not depend on $\sigma_3$. {\bf In other words, as already noticed in \cite{sB99} for lunisolar perturbations, in Eq.~(\ref{eq:ham_SRP}) $\Sigma_3$ is a dummy variable and its value is not important.}

\begin{figure*}
 \includegraphics[width=0.24\textwidth]{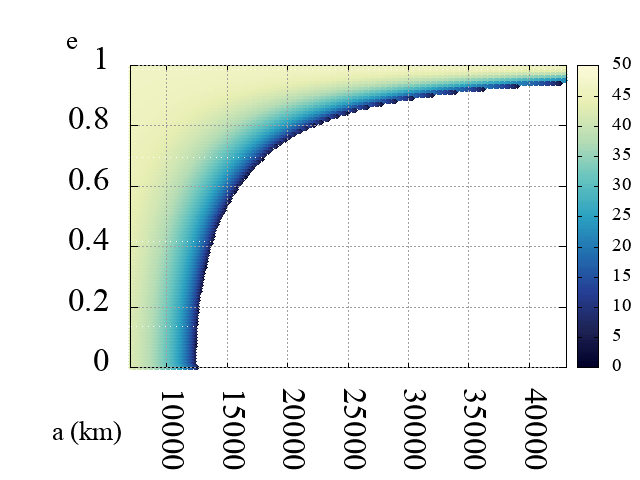} \includegraphics[width=0.24\textwidth]{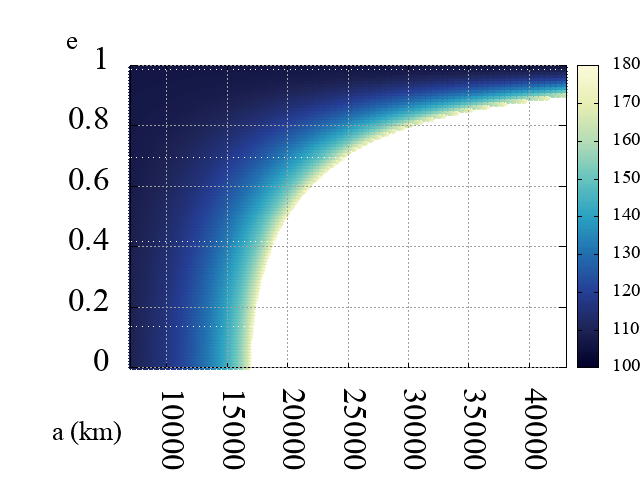}  \includegraphics[width=0.24\textwidth]{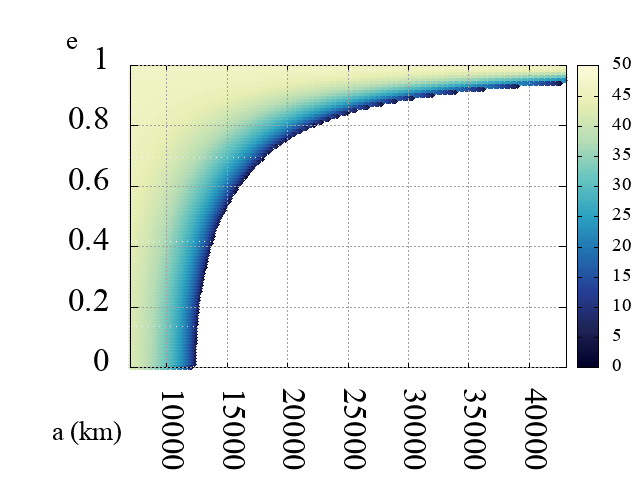} \includegraphics[width=0.24\textwidth]{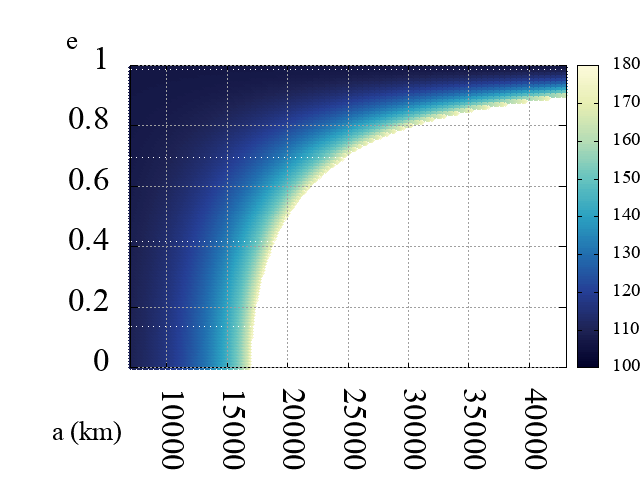}\\
 \includegraphics[width=0.24\textwidth]{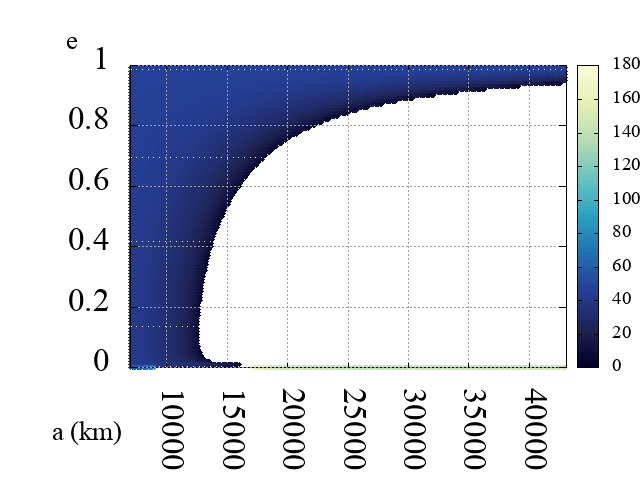} \includegraphics[width=0.24\textwidth]{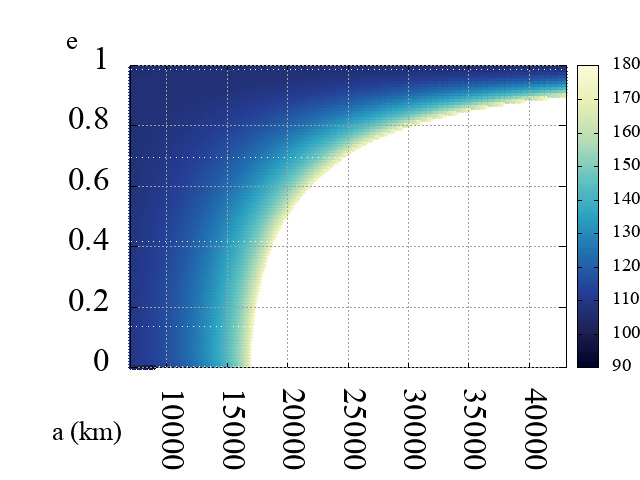}  \includegraphics[width=0.24\textwidth]{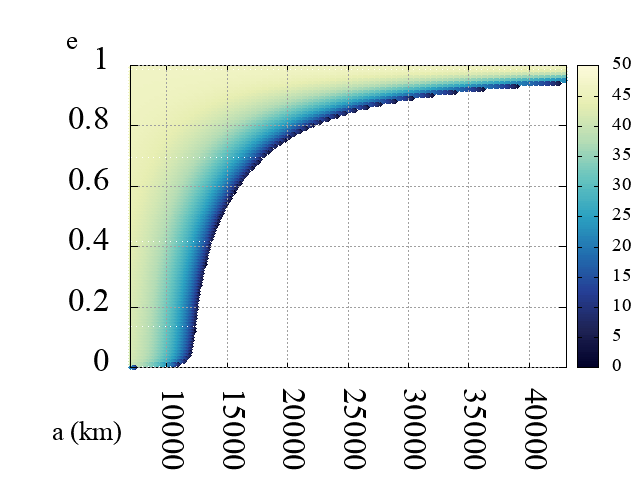} \includegraphics[width=0.24\textwidth]{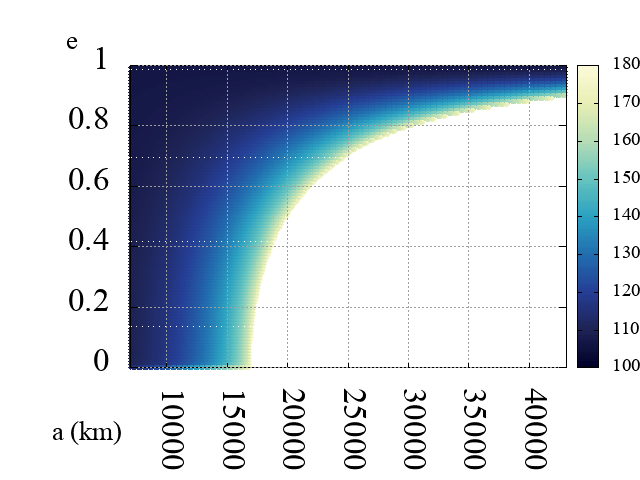}\\
  \includegraphics[width=0.24\textwidth]{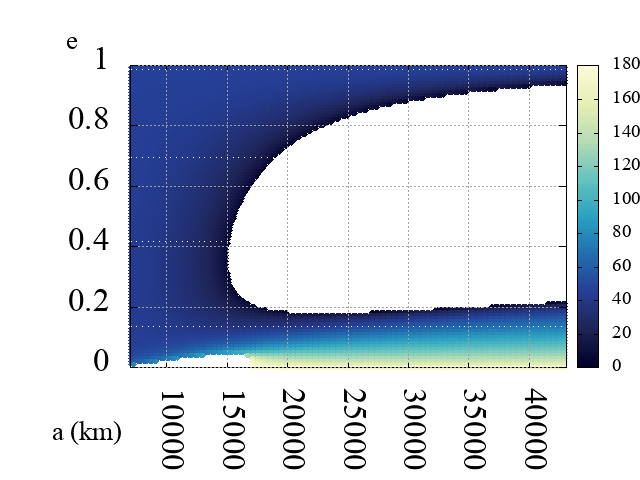} \includegraphics[width=0.24\textwidth]{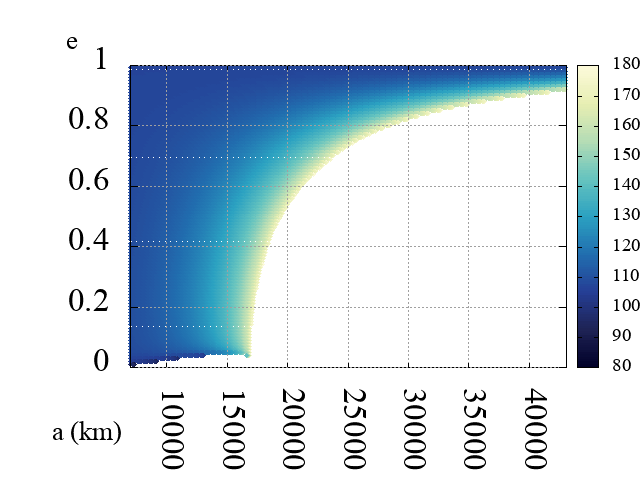}  \includegraphics[width=0.24\textwidth]{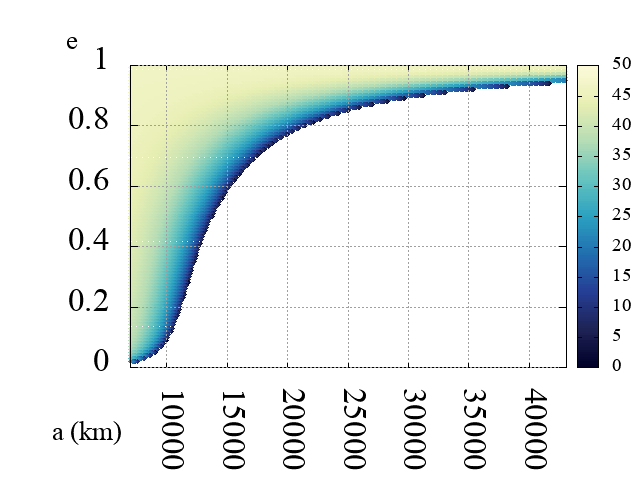} \includegraphics[width=0.24\textwidth]{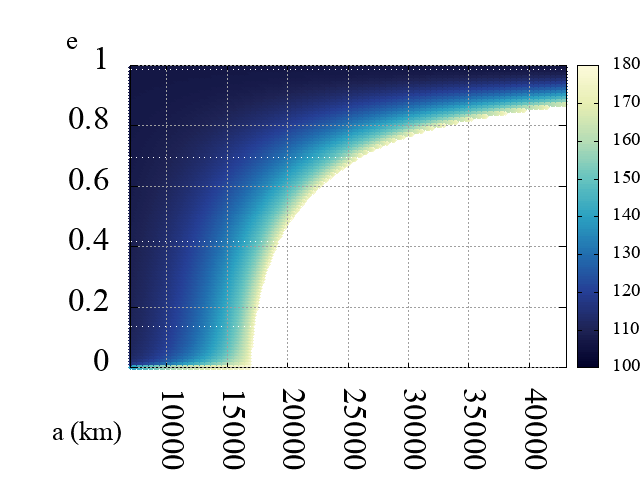}
\caption{As a function of semi-major axis and eccentricity, for $j=1$ we show the inclination solutions (colorbar in degrees) of the quadratic equation, which gives the equilibrium points associated with $\psi=0$ (left two columns), and with $\psi=\pi$ (right two columns). The coefficients of the quadratic equation are given in  Tab.~\ref{tab:quadeqeq0} and Tab.~\ref{tab:quadeqeqpi}, respectively. Top row: $A/m=0.012$ m$^2/$kg; central row: $A/m=1$ m$^2/$kg; bottom row: $A/m=20$ m$^2/$kg.}
\label{fig:aei_res1}       
\end{figure*}

\begin{figure*}
 \includegraphics[width=0.24\textwidth]{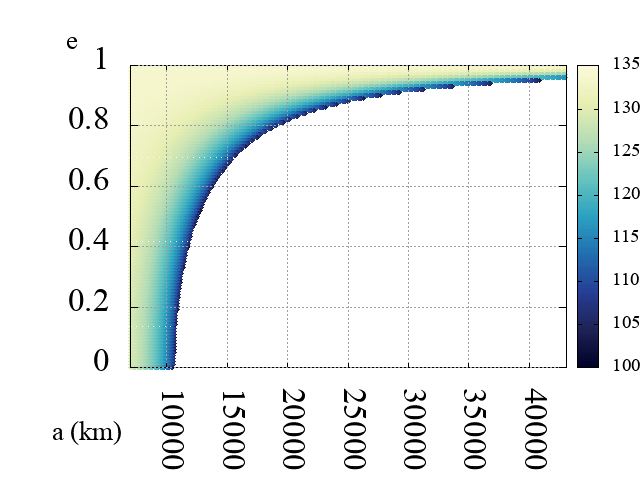} \includegraphics[width=0.24\textwidth]{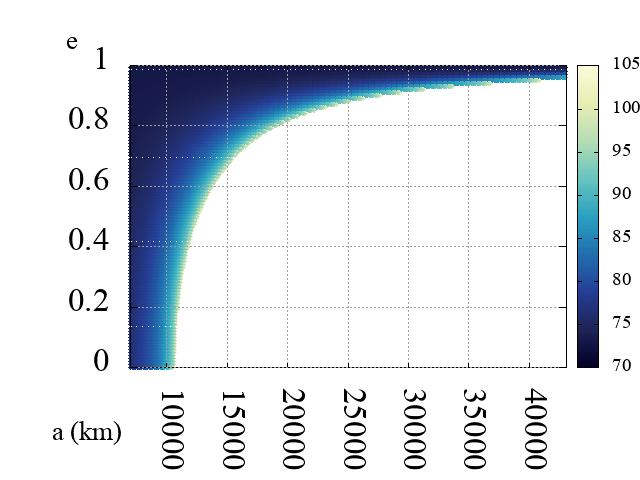}  \includegraphics[width=0.24\textwidth]{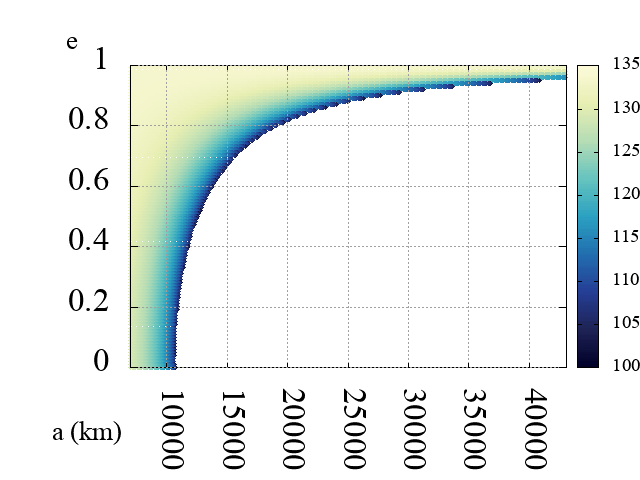} \includegraphics[width=0.24\textwidth]{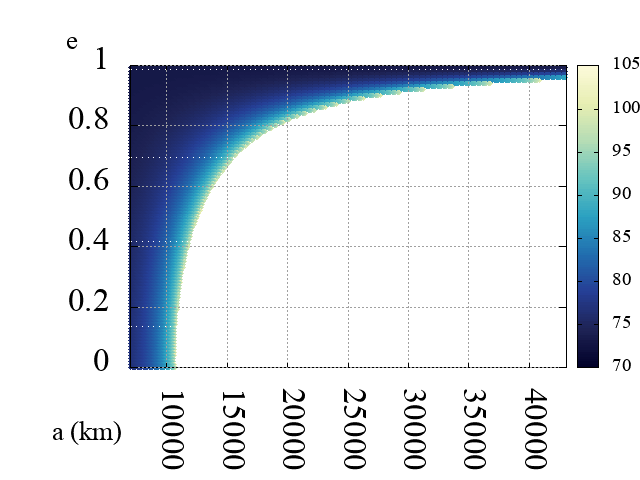}\\
 \includegraphics[width=0.24\textwidth]{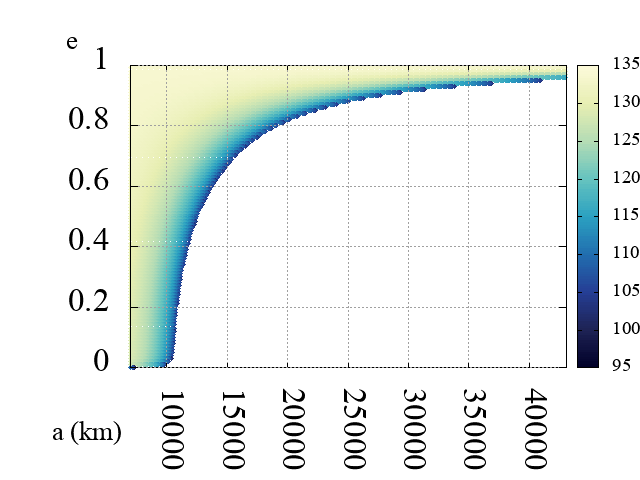} \includegraphics[width=0.24\textwidth]{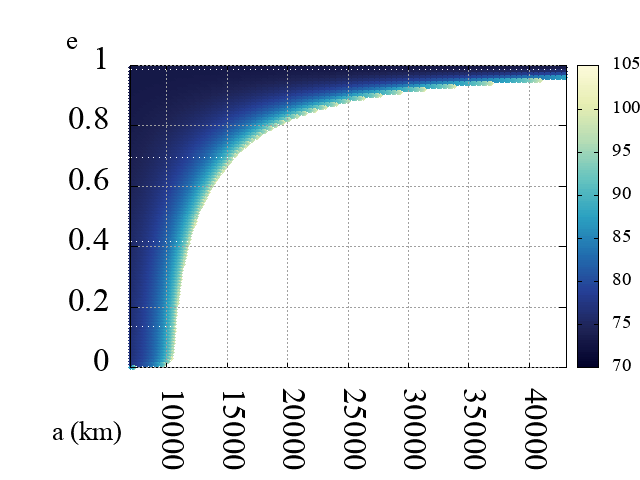}  \includegraphics[width=0.24\textwidth]{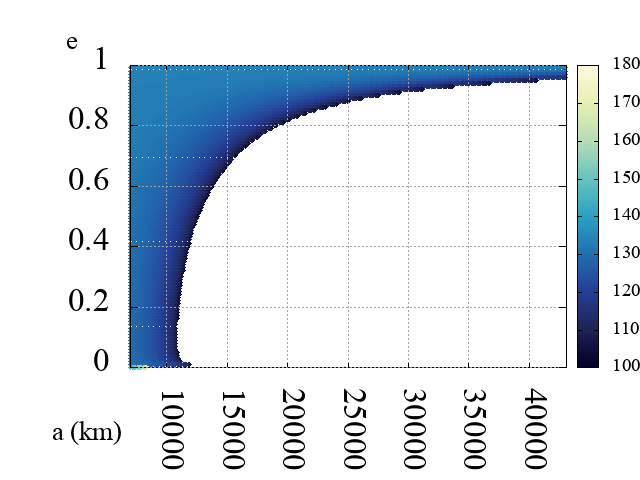} \includegraphics[width=0.24\textwidth]{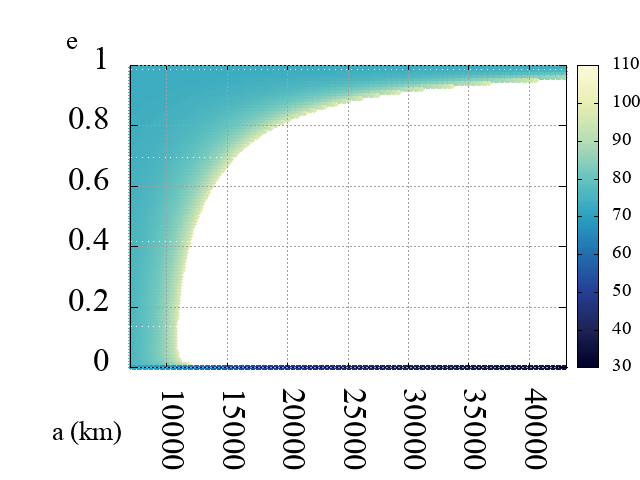}\\
  \includegraphics[width=0.24\textwidth]{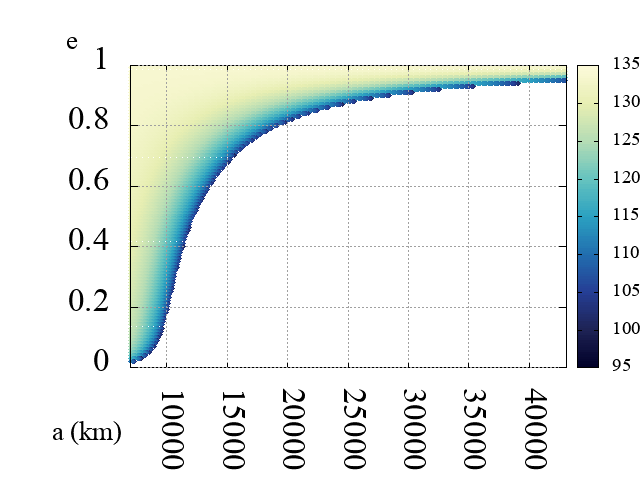} \includegraphics[width=0.24\textwidth]{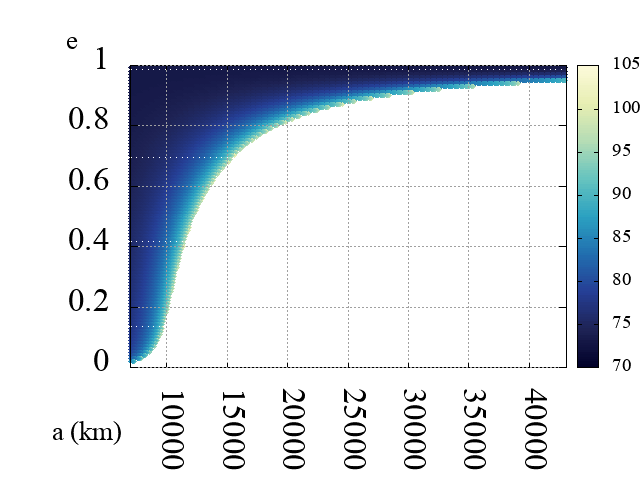}  \includegraphics[width=0.24\textwidth]{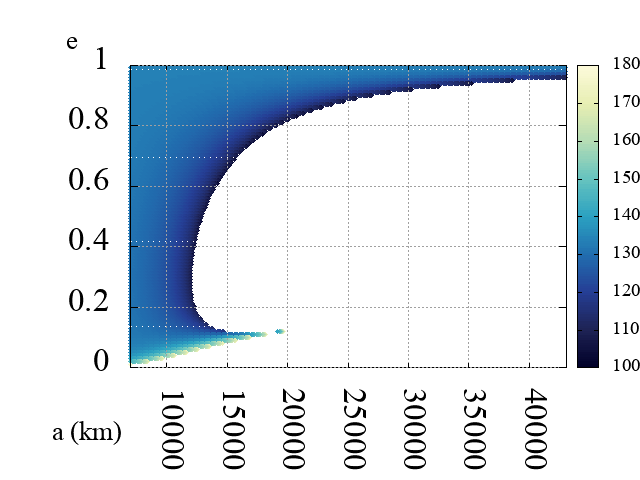} \includegraphics[width=0.24\textwidth]{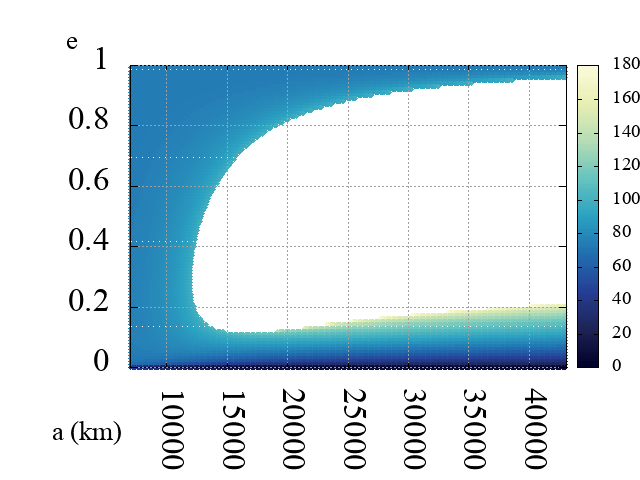}
\caption{The same as in Fig.~\ref{fig:aei_res1}, but for $j=2$.}
\label{fig:aei_res2}       
\end{figure*}

\section{Dynamical Description}
\label{sec:expl}

\begin{figure*}
 \includegraphics[width=0.24\textwidth]{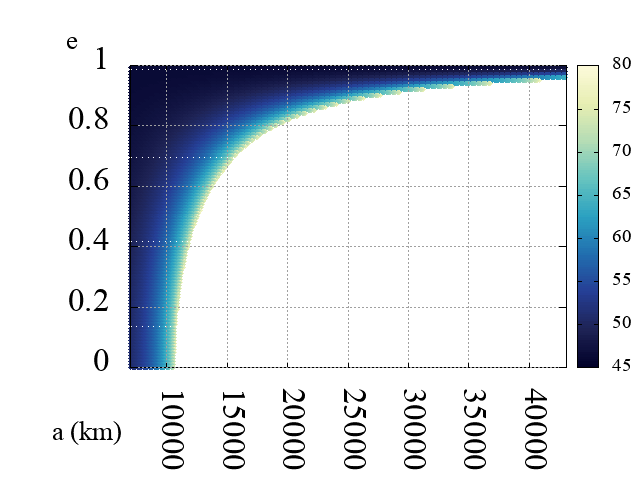} \includegraphics[width=0.24\textwidth]{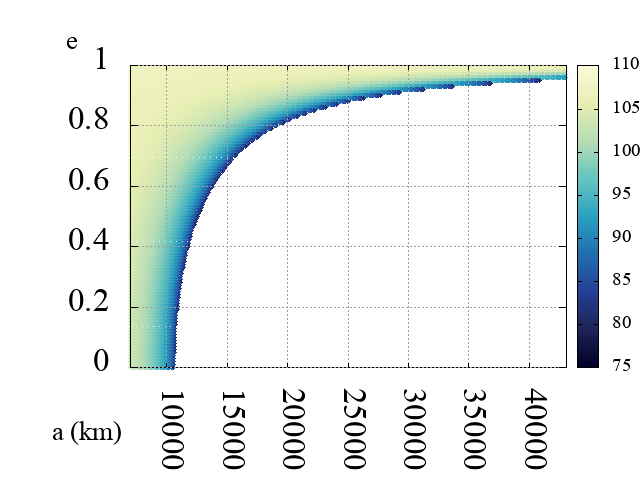}  \includegraphics[width=0.24\textwidth]{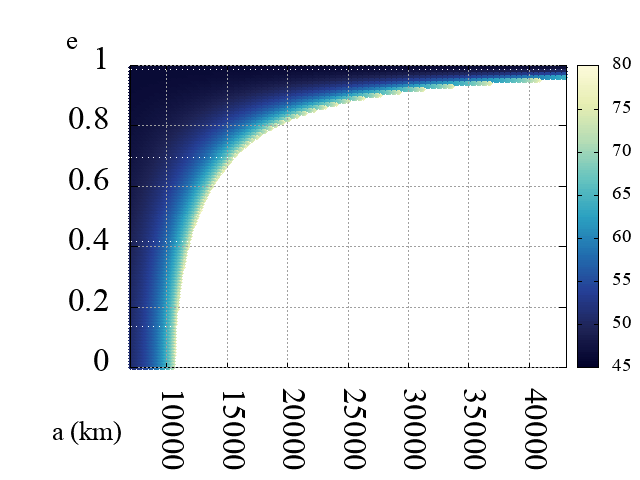} \includegraphics[width=0.24\textwidth]{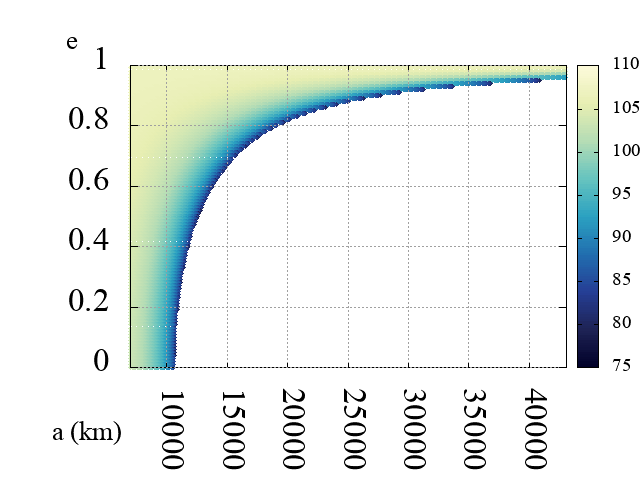}\\
 \includegraphics[width=0.24\textwidth]{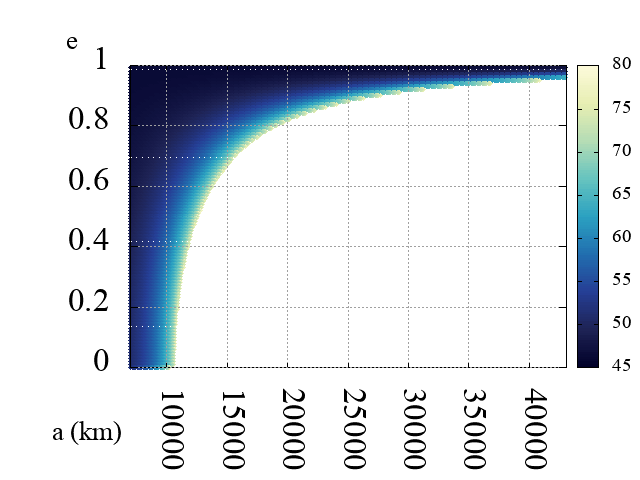} \includegraphics[width=0.24\textwidth]{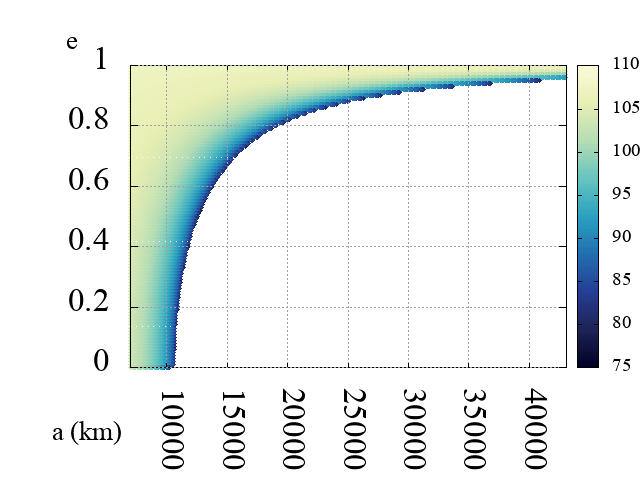}  \includegraphics[width=0.24\textwidth]{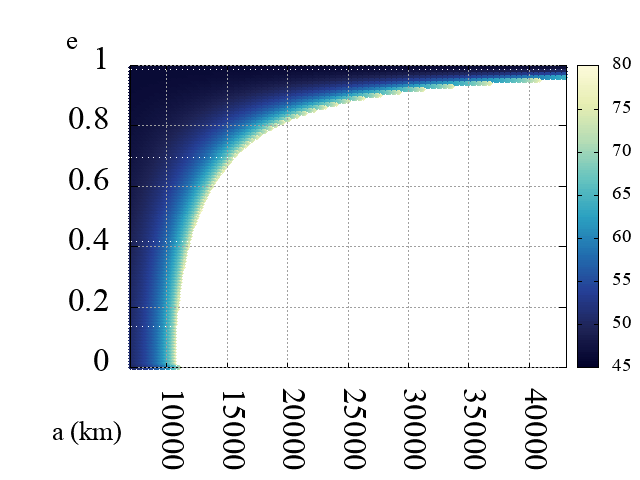} \includegraphics[width=0.24\textwidth]{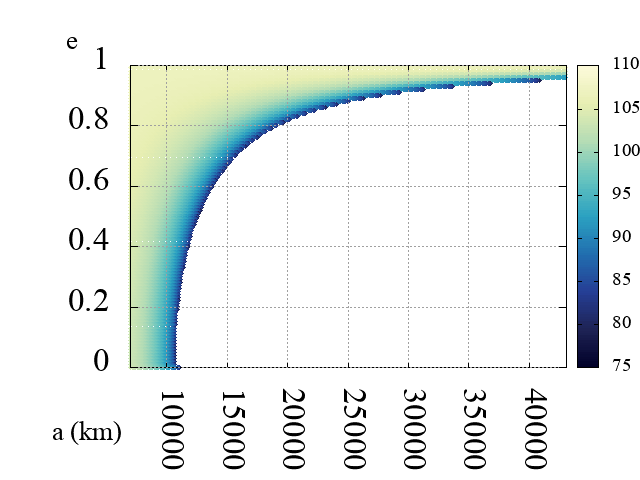}\\
  \includegraphics[width=0.24\textwidth]{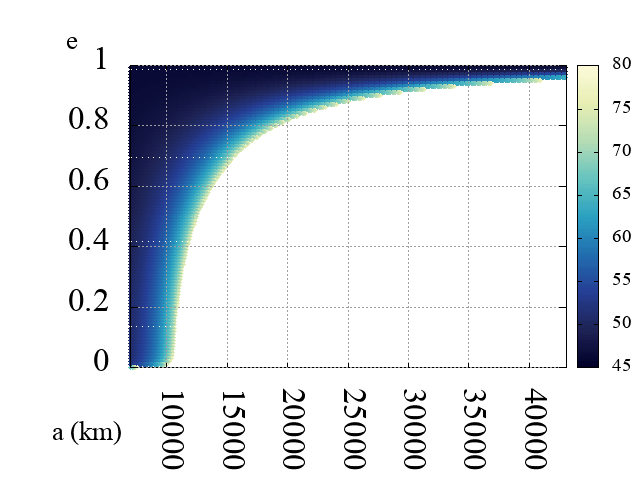} \includegraphics[width=0.24\textwidth]{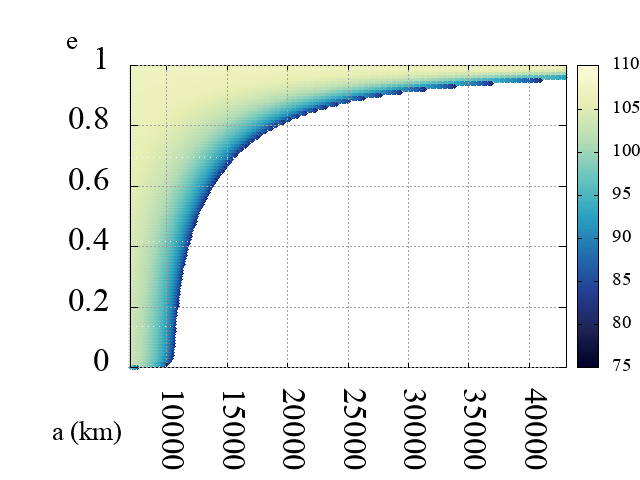}  \includegraphics[width=0.24\textwidth]{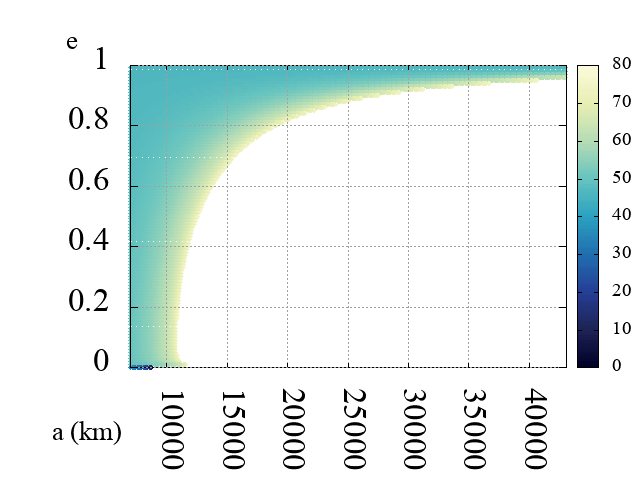} \includegraphics[width=0.24\textwidth]{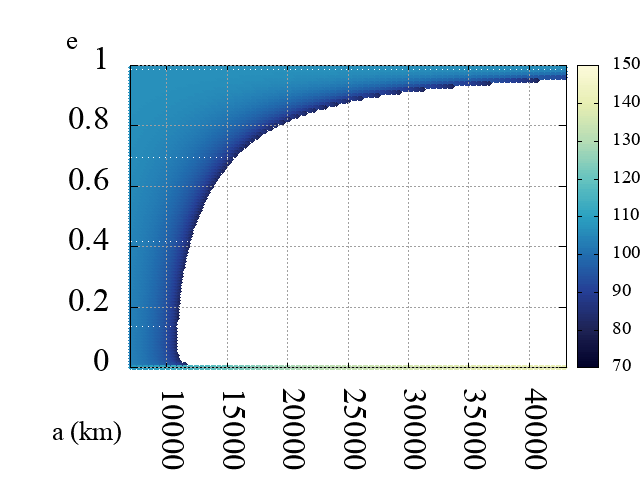}
\caption{The same as in Fig.~\ref{fig:aei_res1}, but for $j=5$.}
\label{fig:aei_res5}       
\end{figure*}

\begin{figure*}
 \includegraphics[width=0.24\textwidth]{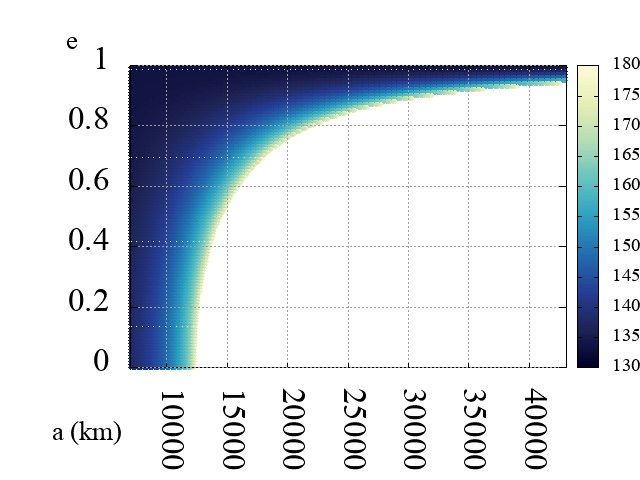} \includegraphics[width=0.24\textwidth]{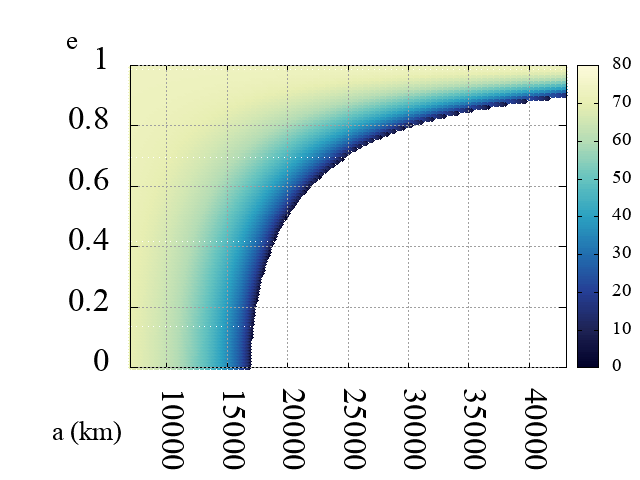}  \includegraphics[width=0.24\textwidth]{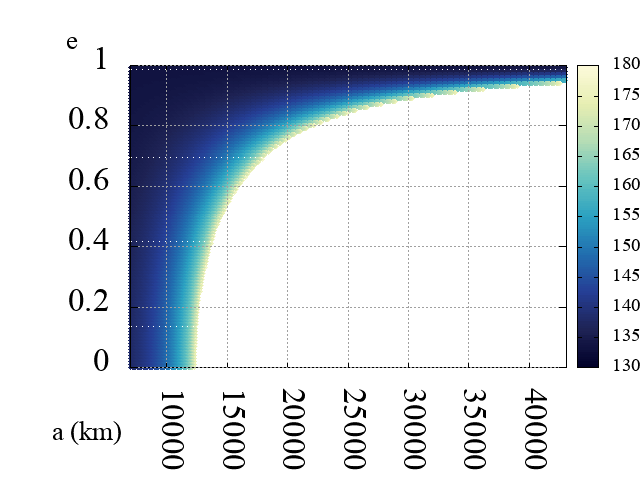} \includegraphics[width=0.24\textwidth]{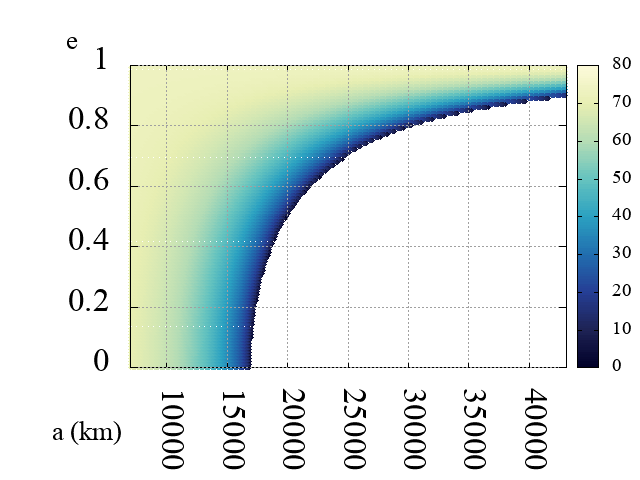}\\
 \includegraphics[width=0.24\textwidth]{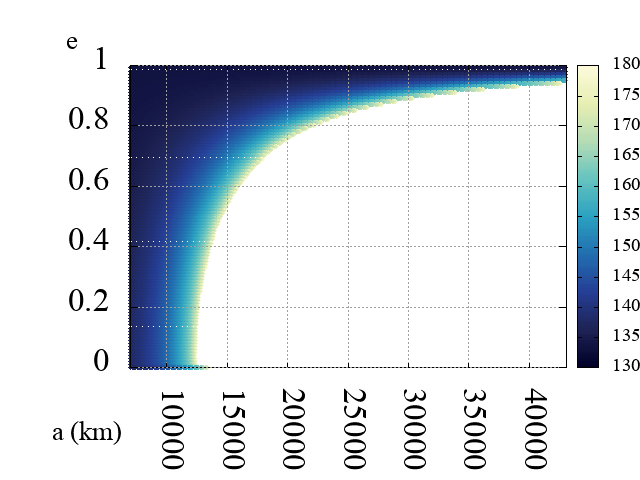} \includegraphics[width=0.24\textwidth]{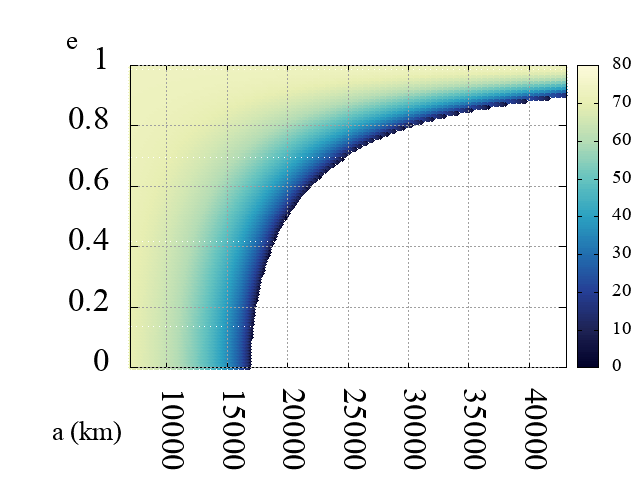}  \includegraphics[width=0.24\textwidth]{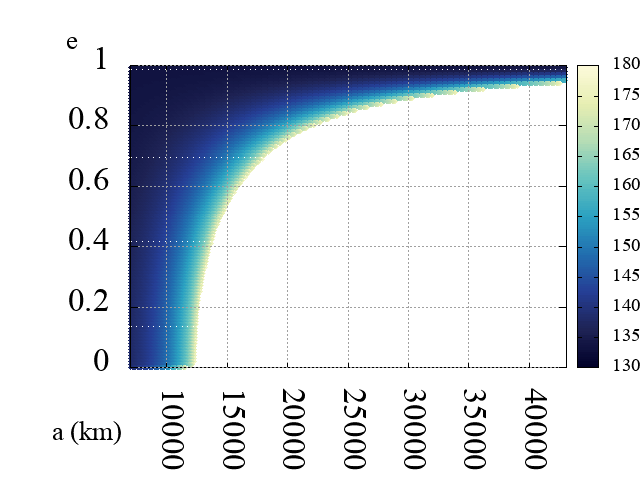} \includegraphics[width=0.24\textwidth]{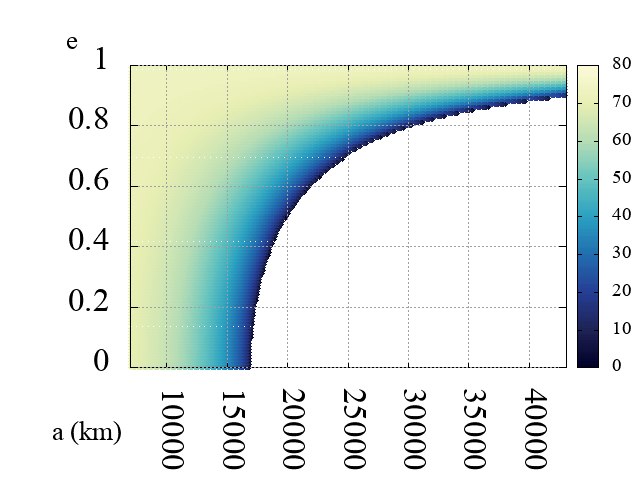}\\
  \includegraphics[width=0.24\textwidth]{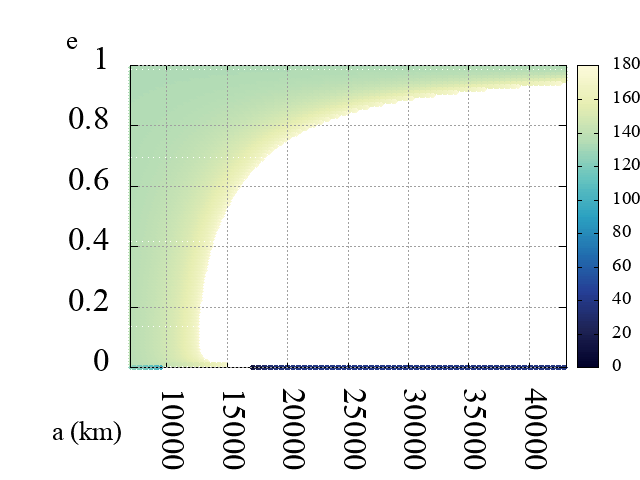} \includegraphics[width=0.24\textwidth]{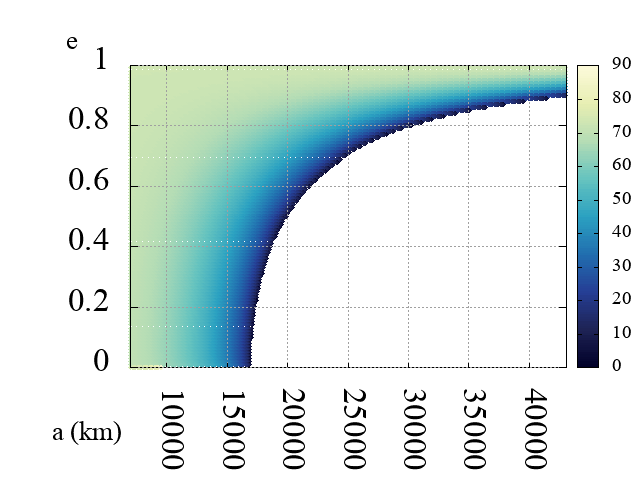}  \includegraphics[width=0.24\textwidth]{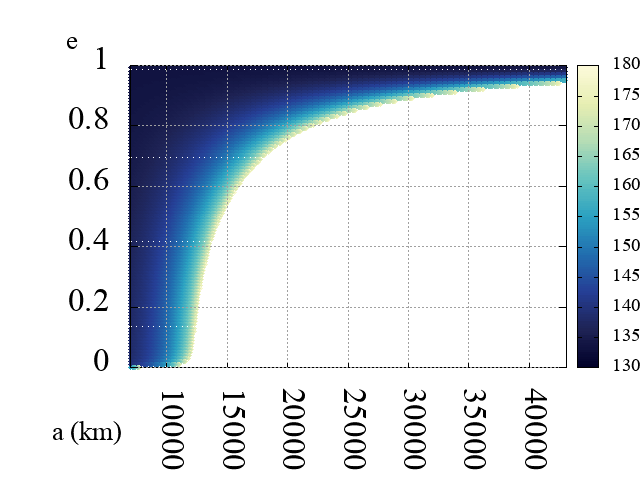} \includegraphics[width=0.24\textwidth]{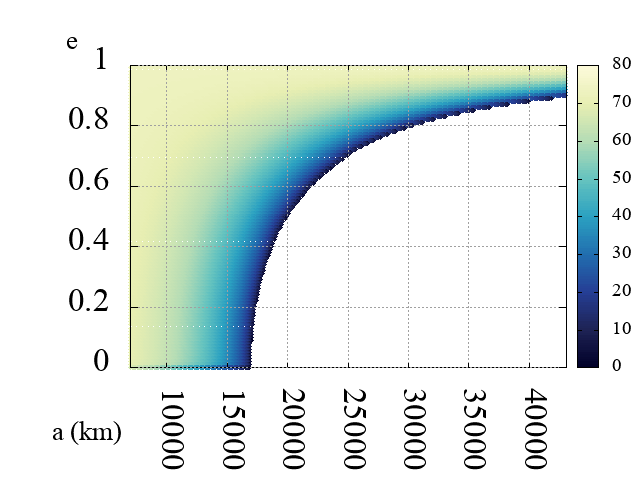}
\caption{The same as in Fig.~\ref{fig:aei_res1}, but for $j=6$.}
\label{fig:aei_res6}       
\end{figure*}

In Figs.~\ref{fig:aei_res1}-\ref{fig:aei_res6}, we show the location of the equilibrium points at $\psi_j=0$ and $\psi_j=\pi$, associated with each term $j=1,2,5,6$, as a function of $(a,e,i)$ for three values of area-to-mass ratio. These are $A/m=0.012$ m$^2/$kg, $A/m=1$ m$^2/$kg and $A/m=20$ m$^2/$kg, characteristic of a standard satellite, a satellite equipped with an area-augmentation device with the present technology \cite{REF3} and a high area-to-mass ratio debris fragment, respectively. 

{\bf The solutions corresponding to inclination equal to zero, for $j=1$ and $j=2$, are equivalent to the ones described in \cite{REF1}. For $A/m=0.012$ m$^2/$kg and  $j=1$ or $j=2$, the solutions presented in \cite{REF2}, for the inclined case when considering only $J_2$, are equivalent to the ones shown here, because for low $A/m$ values the effect of the solar radiation pressure on $(\Omega, \omega)$ is negligible.} 

\begin{figure*}
 \includegraphics[width=0.24\textwidth]{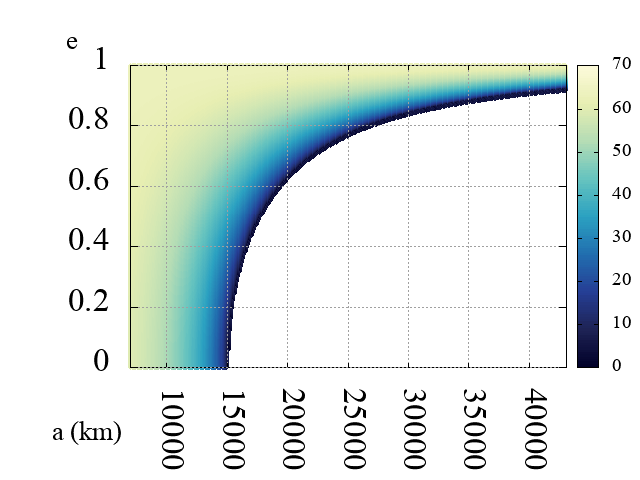} \includegraphics[width=0.24\textwidth]{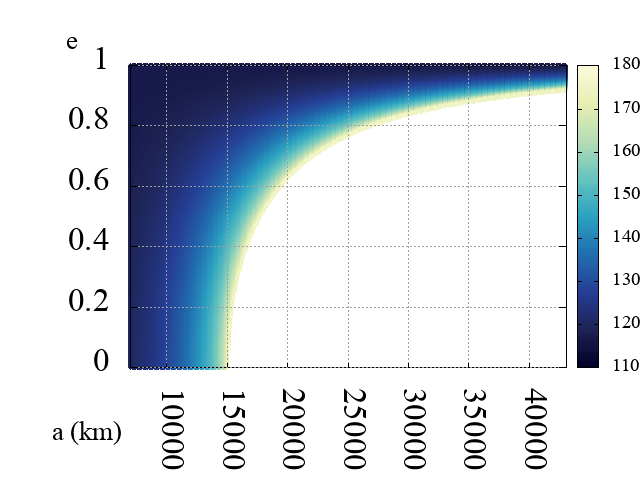}  \includegraphics[width=0.24\textwidth]{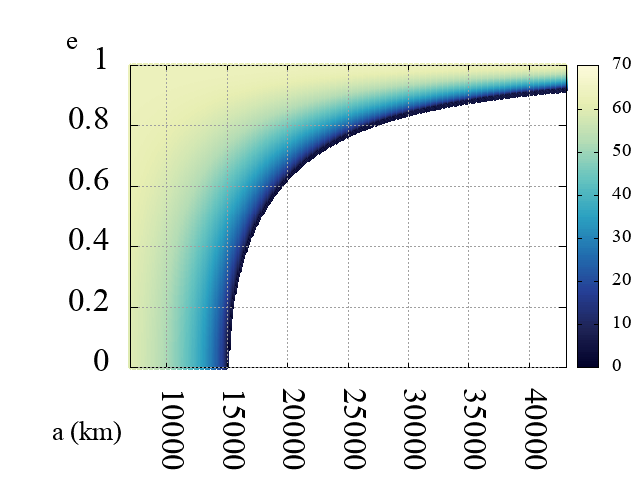} \includegraphics[width=0.24\textwidth]{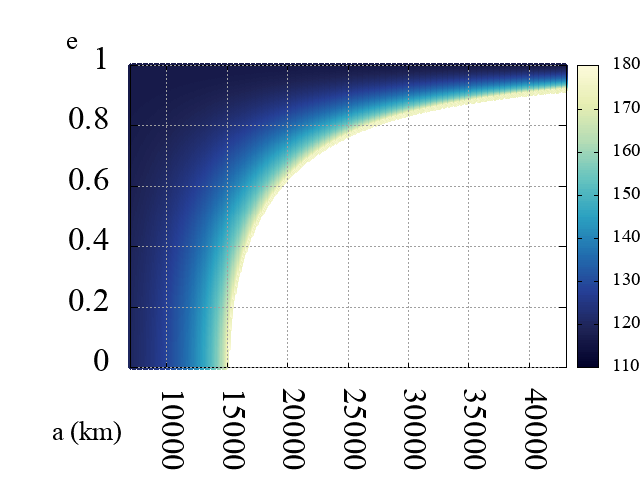}\\
  \includegraphics[width=0.24\textwidth]{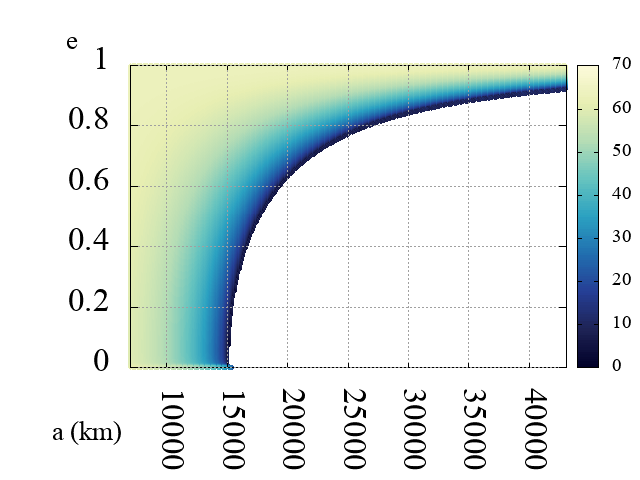} \includegraphics[width=0.24\textwidth]{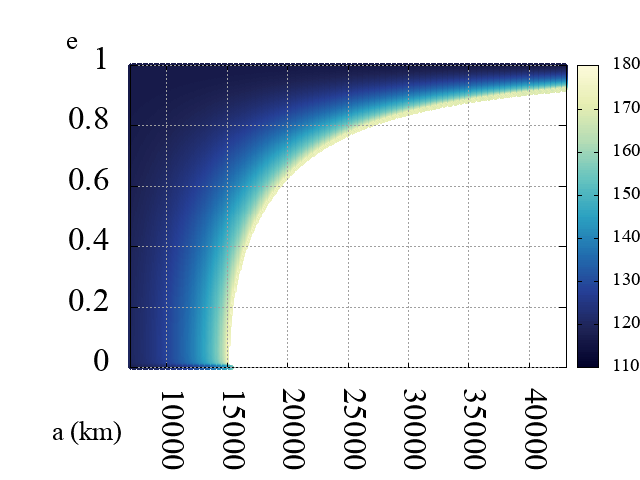}  \includegraphics[width=0.24\textwidth]{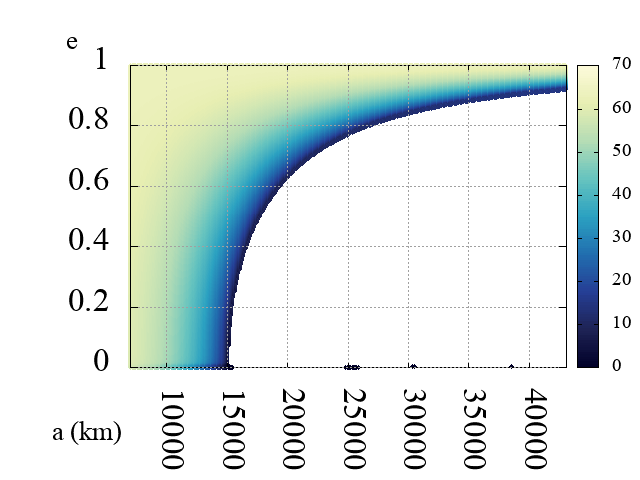} \includegraphics[width=0.24\textwidth]{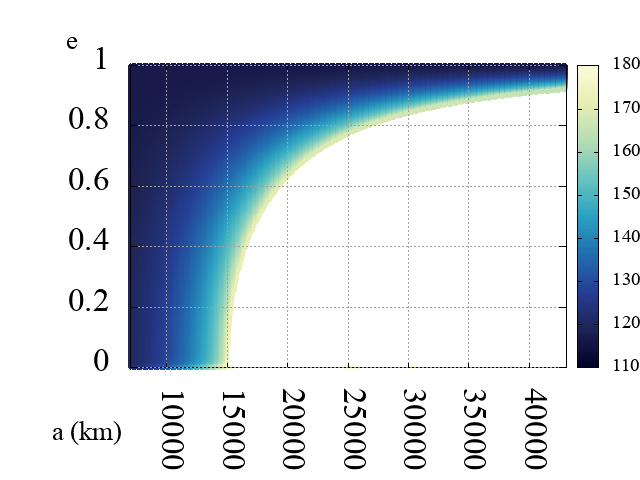}\\
  \includegraphics[width=0.24\textwidth]{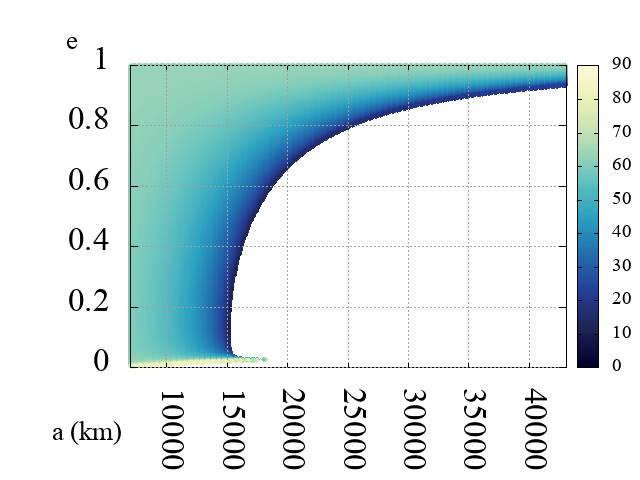} \includegraphics[width=0.24\textwidth]{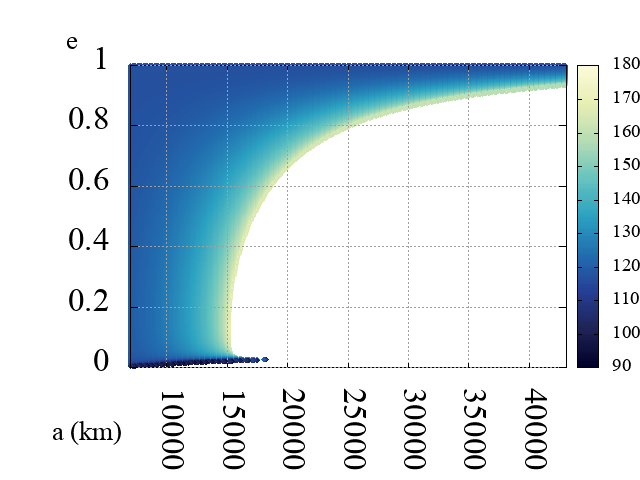}  \includegraphics[width=0.24\textwidth]{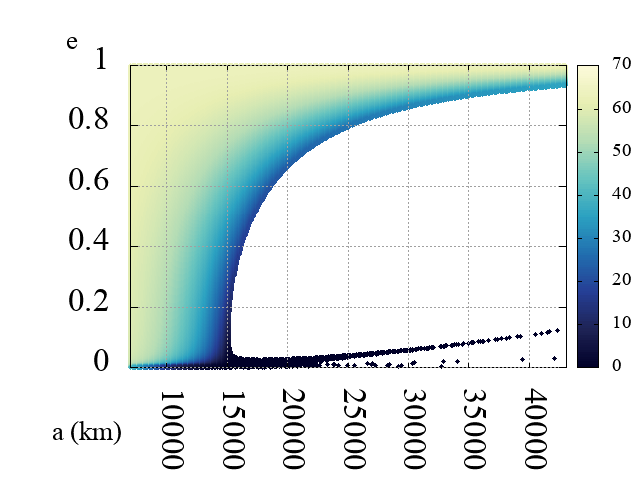} \includegraphics[width=0.24\textwidth]{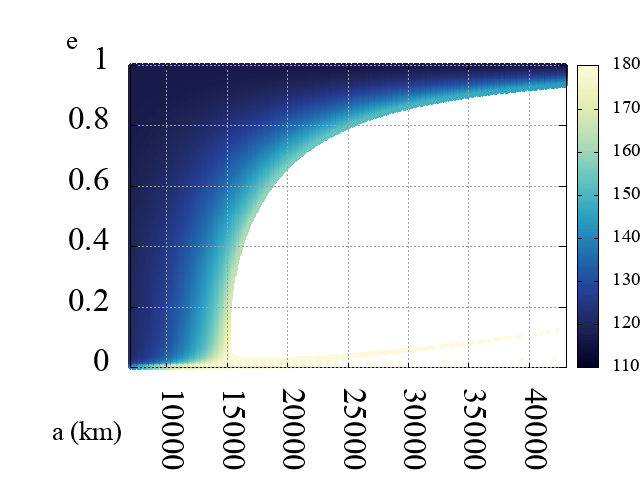}
\caption{As a function of semi-major axis and eccentricity, for $j=3$ we show the inclination solutions (colorbar in degrees) of the cubic equation, which gives the equilibrium points associated with $\psi=0$ (left two columns), and with $\psi=\pi$ (right two columns). The coefficients of the cubic equation are given in  Tab.~\ref{tab:cubeqeq0} and Tab.~\ref{tab:cubeqeqpi}, respectively. Top row: $A/m=0.012$ m$^2/$kg; central row: $A/m=1$ m$^2/$kg; bottom row: $A/m=20$ m$^2/$kg.}
\label{fig:aei_res3}       
\end{figure*}

\begin{figure*}
 \includegraphics[width=0.24\textwidth]{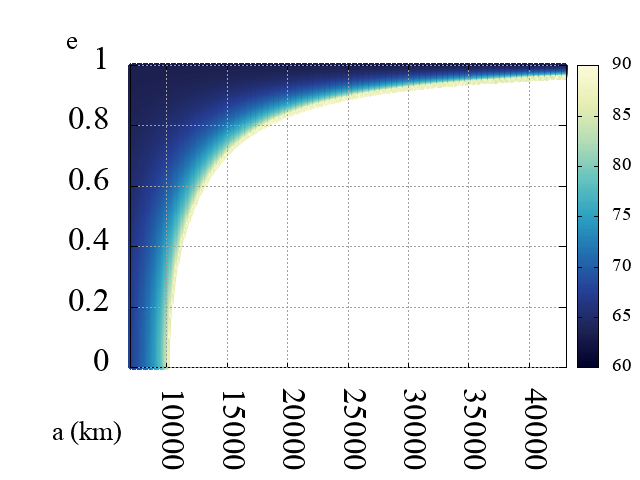} \includegraphics[width=0.24\textwidth]{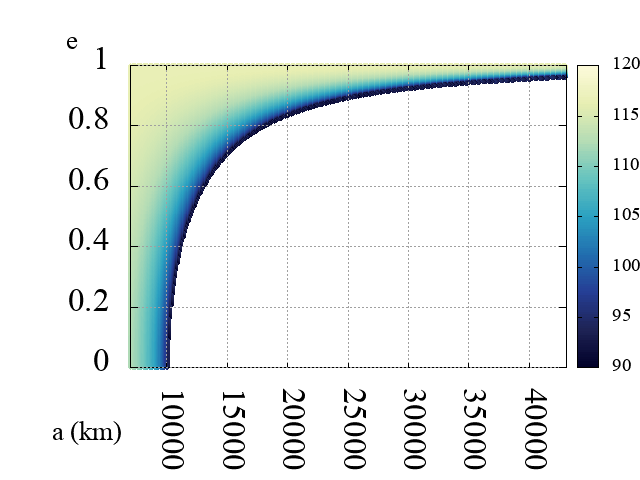}  \includegraphics[width=0.24\textwidth]{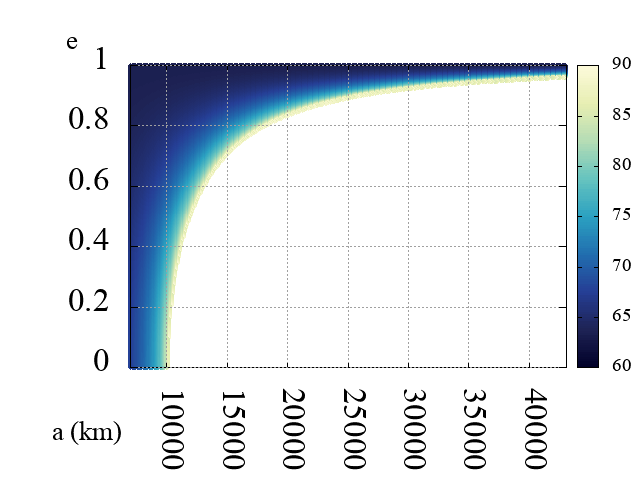} \includegraphics[width=0.24\textwidth]{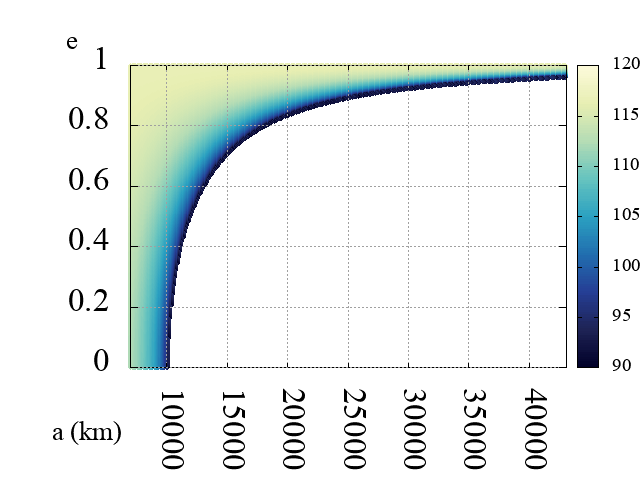}\\
  \includegraphics[width=0.24\textwidth]{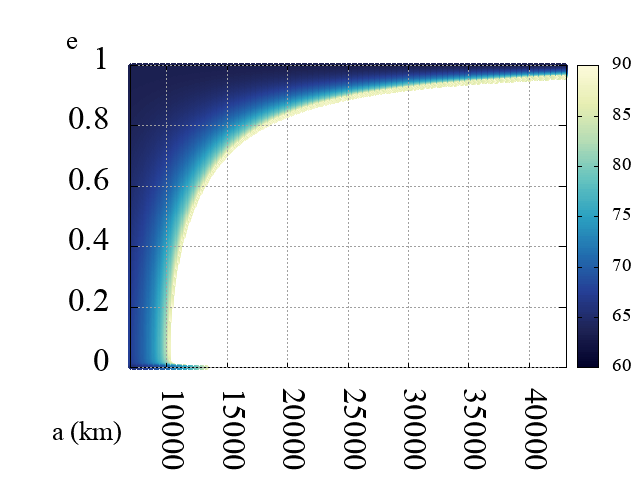} \includegraphics[width=0.24\textwidth]{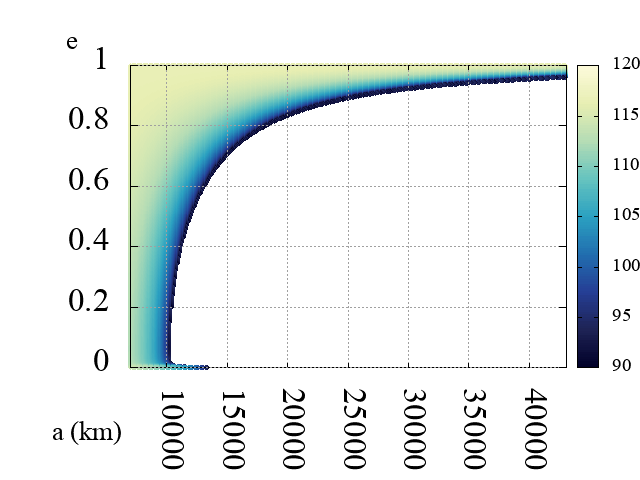}  \includegraphics[width=0.24\textwidth]{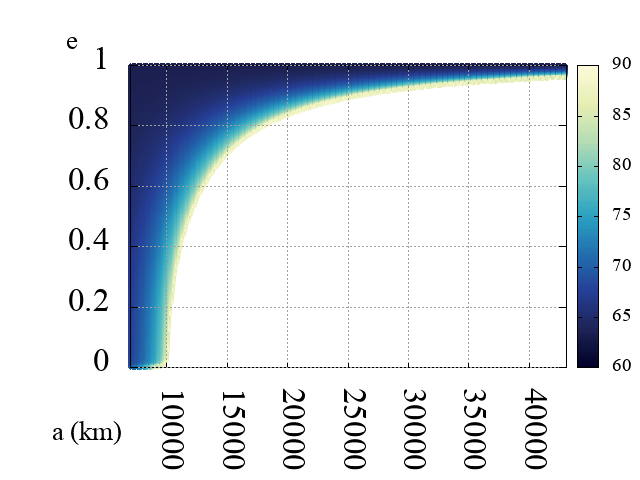} \includegraphics[width=0.24\textwidth]{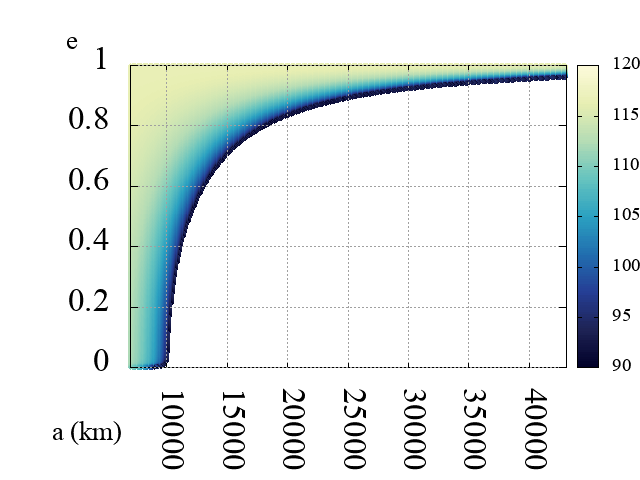}\\
  \includegraphics[width=0.24\textwidth]{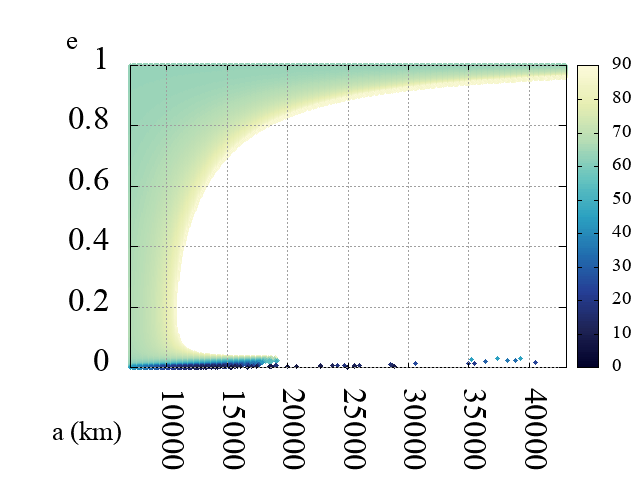} \includegraphics[width=0.24\textwidth]{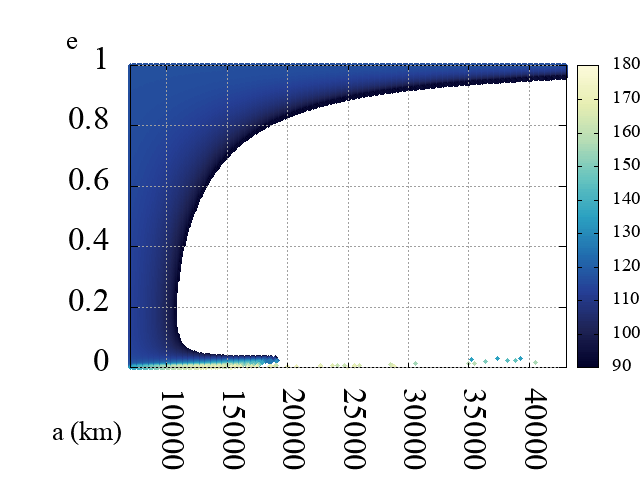}  \includegraphics[width=0.24\textwidth]{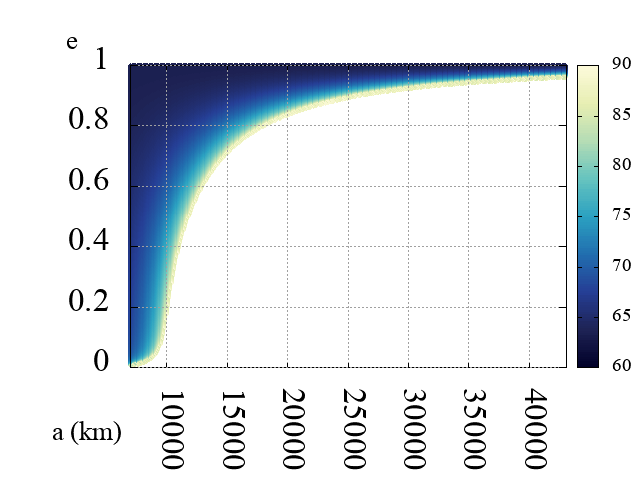} \includegraphics[width=0.24\textwidth]{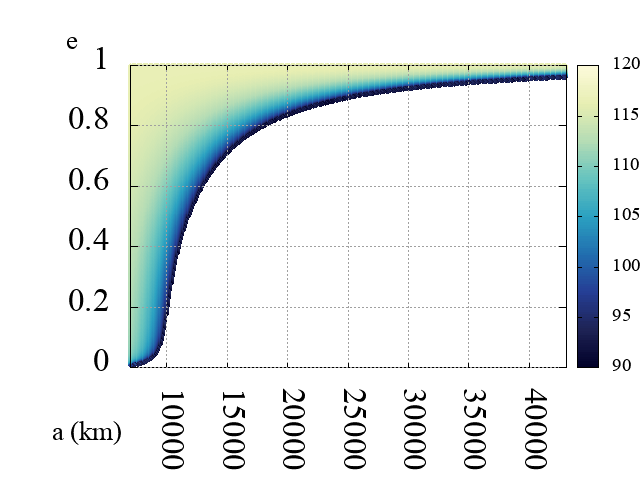}
\caption{The same as in Fig.~\ref{fig:aei_res3}, but for $j=4$.}
\label{fig:aei_res4}       
\end{figure*}

Notice how the inclination changes as the semi-major axis increases, increasing or decreasing depending on $j$ and on whether the equilibrium is at $\psi_j=0$ or $\psi_j=\pi$. Note also that, by increasing the area-to-mass ratio, the main effect is to enlarge the range of semi-major axis where the effect can be detected. In general, for quasi-circular orbits, the dynamics is coupled up to about $a=15000$ km, except for the cases  $\psi_1=0$ and  $\psi_2=\pi$ of the debris fragment, for which the effect can be considerable up to Geosynchronous Earth Orbits (GEO) altitudes. In Figs.~\ref{fig:aei_res3}-\ref{fig:aei_res4}, we show the inclination values corresponding to the equilibrium points for $j=3$ and $j=4$, as a function of $(a,e)$ for the same three values of area-to-mass ratio. The inclination showed is computed by considering one of the solutions of the cubic equations displayed in Tabs.~\ref{tab:cubeqeq0}-\ref{tab:cubeqeqpi}. The other solutions correspond to the opposite value showed and to the singularity $i=0.$

The dynamics under consideration is organized according to the equilibrium points which exist for a given $(a,\Lambda,A/m,j)$ combination.  In other words, to obtain a single portrait of the behavior of the phase space, the information displayed in the previous figures has to be accompanied by the information on $\Lambda$. In particular, given $(a,A/m,j)$, it is possible to distinguish regions in terms of number and type of equilibrium points, by computing the minimum and maximum values attained by the curves, which represent the evolution of $\Lambda$, as a function of $e$, at $\psi_j=0$ and $\psi_j=\pi$. The type of equilibrium point is determined by its linear behavior: from the analysis performed for this work for orbits around the Earth (from LEO to GEO altitudes), the point can be either a center or a saddle. Note that $A/m$ is constant as given by the problem, $a$ is constant under the effects considered (see Eqs.~(\ref{eq:equmot})), while the dominant resonant term $j$ might change along a curve, as it will be showed in the following. 

\begin{figure*}
 \includegraphics[width=0.49\textwidth]{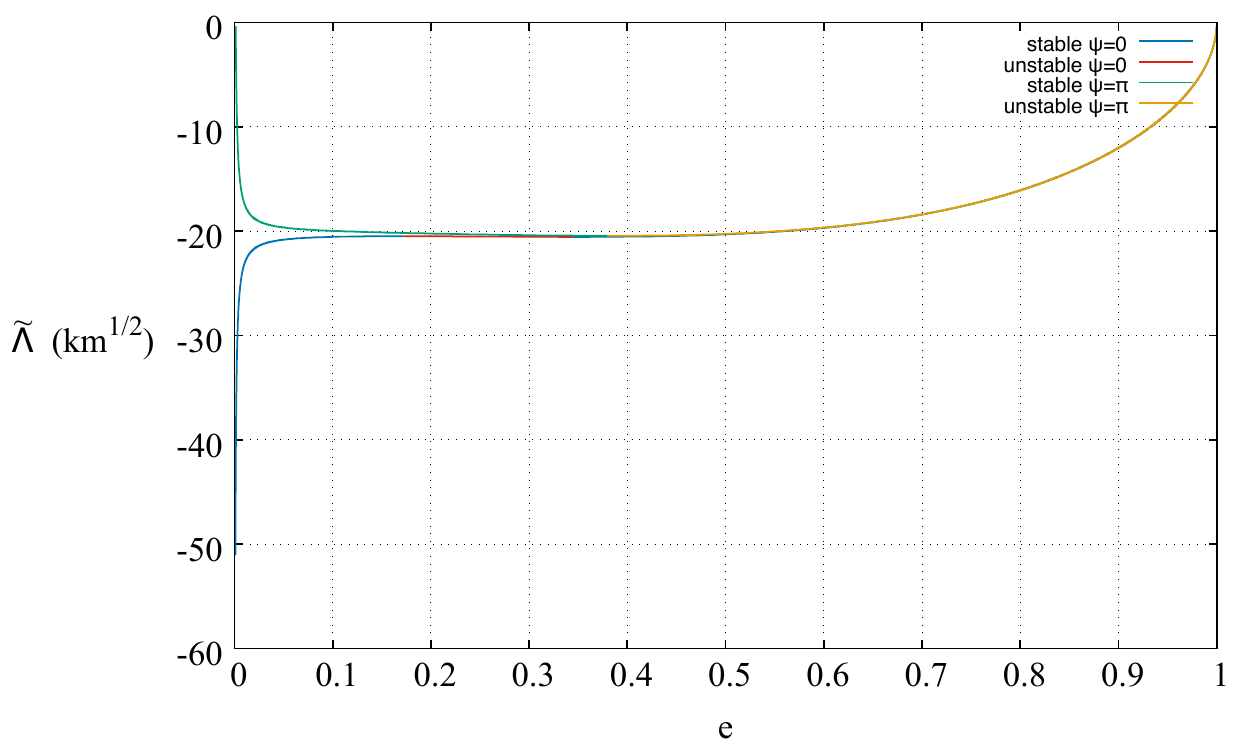}  \includegraphics[width=0.49\textwidth]{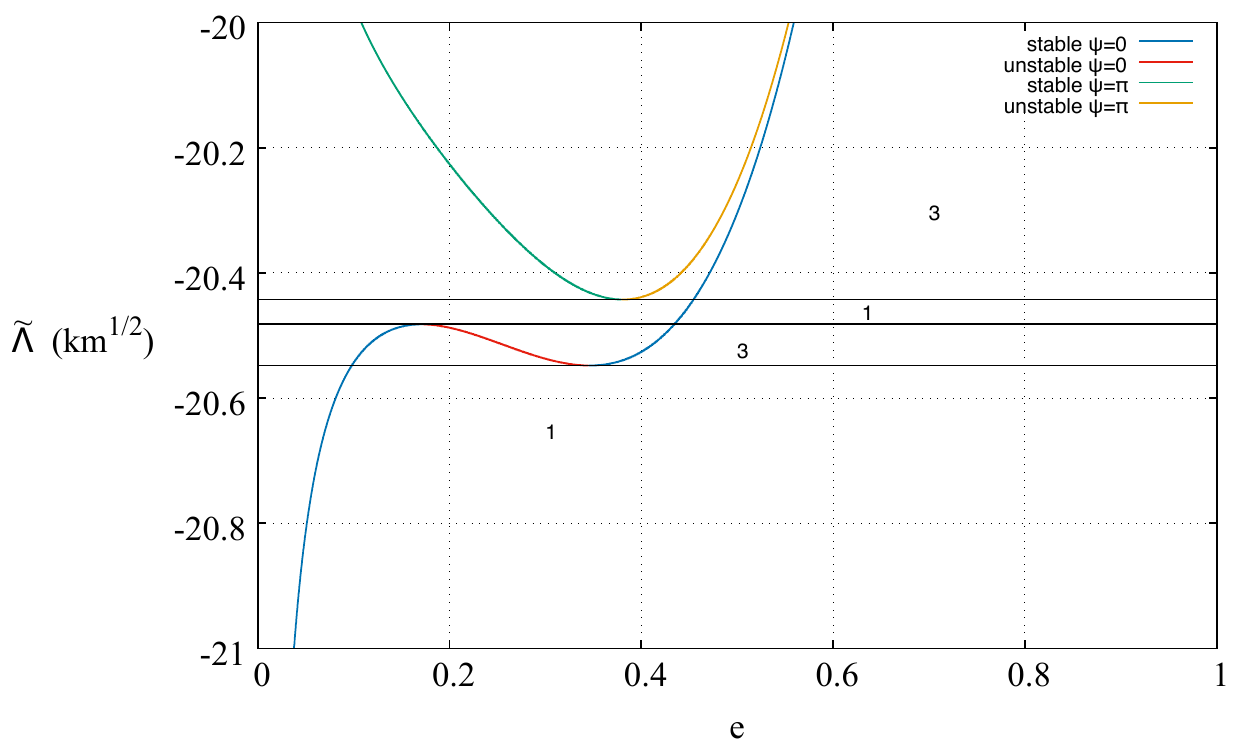} 
\caption{$\tilde{\Lambda}$ as a function of eccentricity, at the equilibrium points existing for $j=1$, $A/m=1$ m$^2/$kg and $a=8078$ km. On the right a close-up view, showing also the number of equilibrium points in the different regions. Blue: stable point at $\psi_1=0$; red: unstable point at $\psi_1=0$. Green: stable point at $\psi_1=\pi$; ochre: unstable point at $\psi_1=\pi$.}
\label{fig:esempio_Lambdae_res1_8078_a2m1}       
\end{figure*}

\begin{figure*}
\centering
 \includegraphics[width=0.49\textwidth]{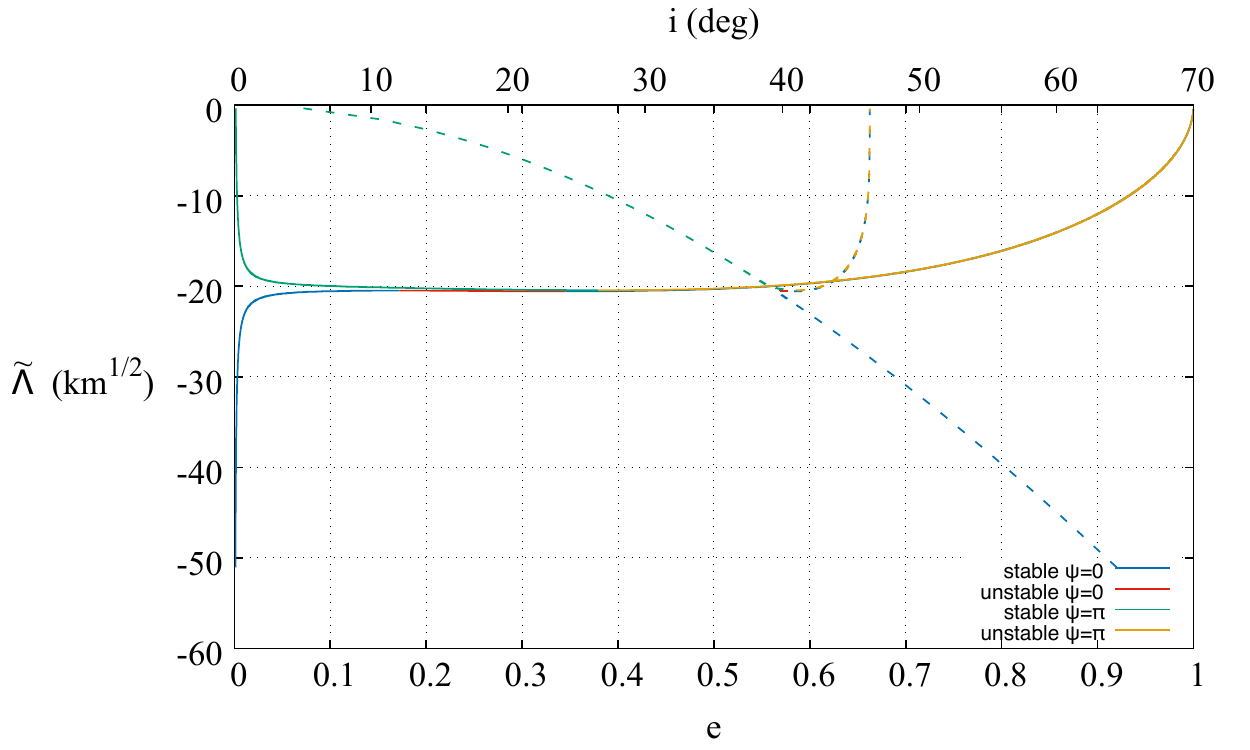}  \includegraphics[width=0.49\textwidth]{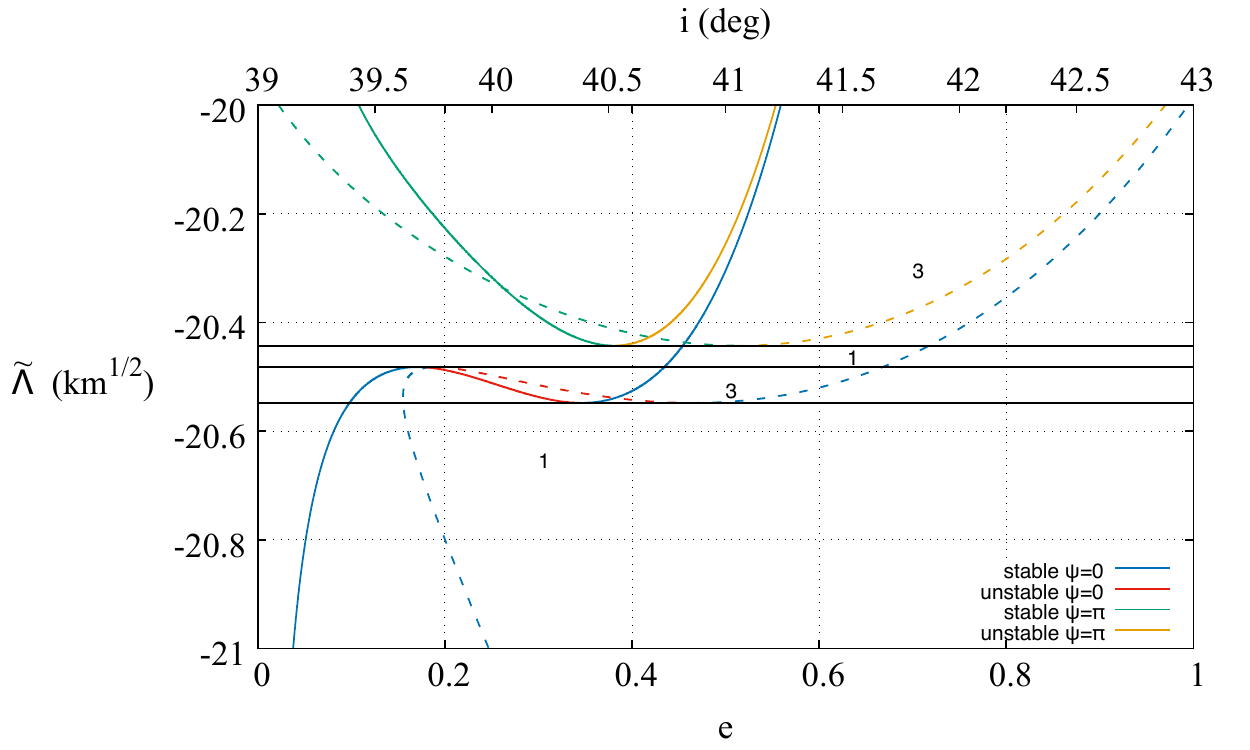} 
\caption{$\tilde{\Lambda}$ as a function of eccentricity, at the equilibrium points existing for $j=1$, $A/m=1$ m$^2/$kg and $a=8078$ km. On the right a close-up view. Blue: stable point at $\psi_1=0$; red: unstable point at $\psi_1=0$. Green: stable point at $\psi_1=\pi$; ochre: unstable point at $\psi_1=\pi$. The dashed curves show the value of inclination corresponding to the given $\Lambda$, the color code is the same.}
\label{fig:esempio_Lambdaei_res1_8078_a2m1}       
\end{figure*}

\begin{figure*}
\centering
 \includegraphics[width=0.49\textwidth]{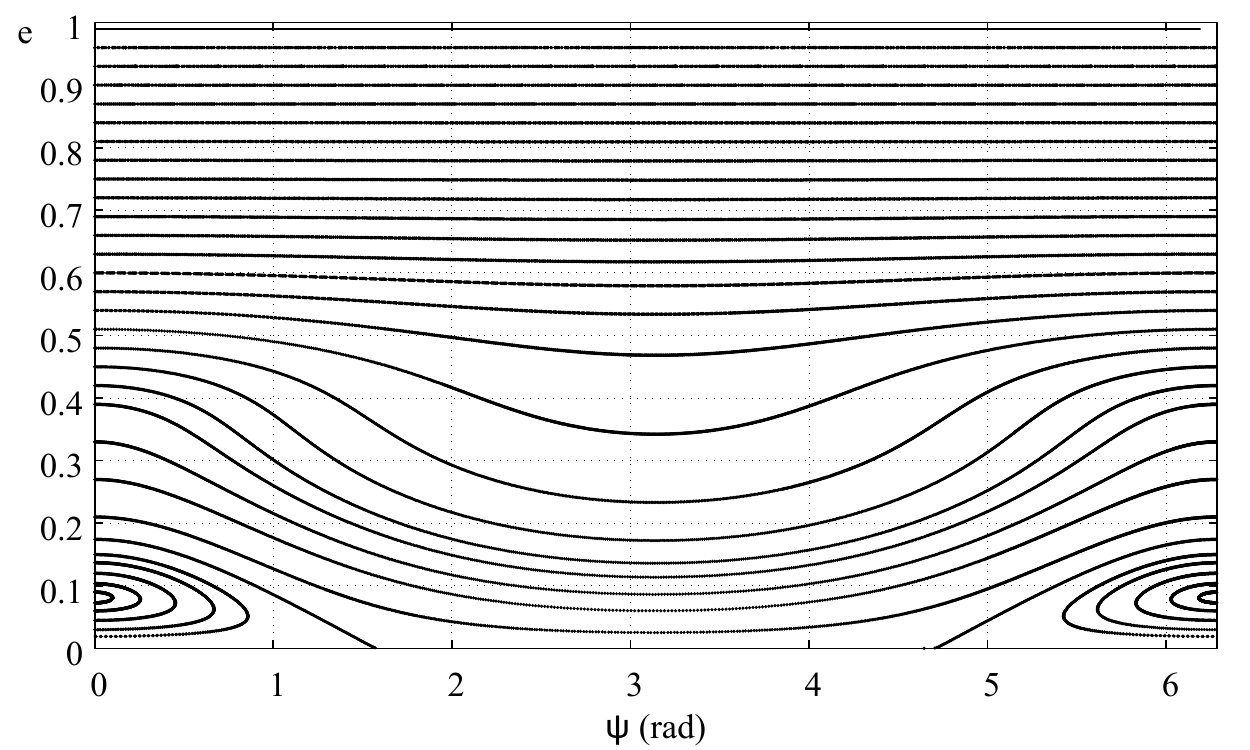}   \includegraphics[width=0.49\textwidth]{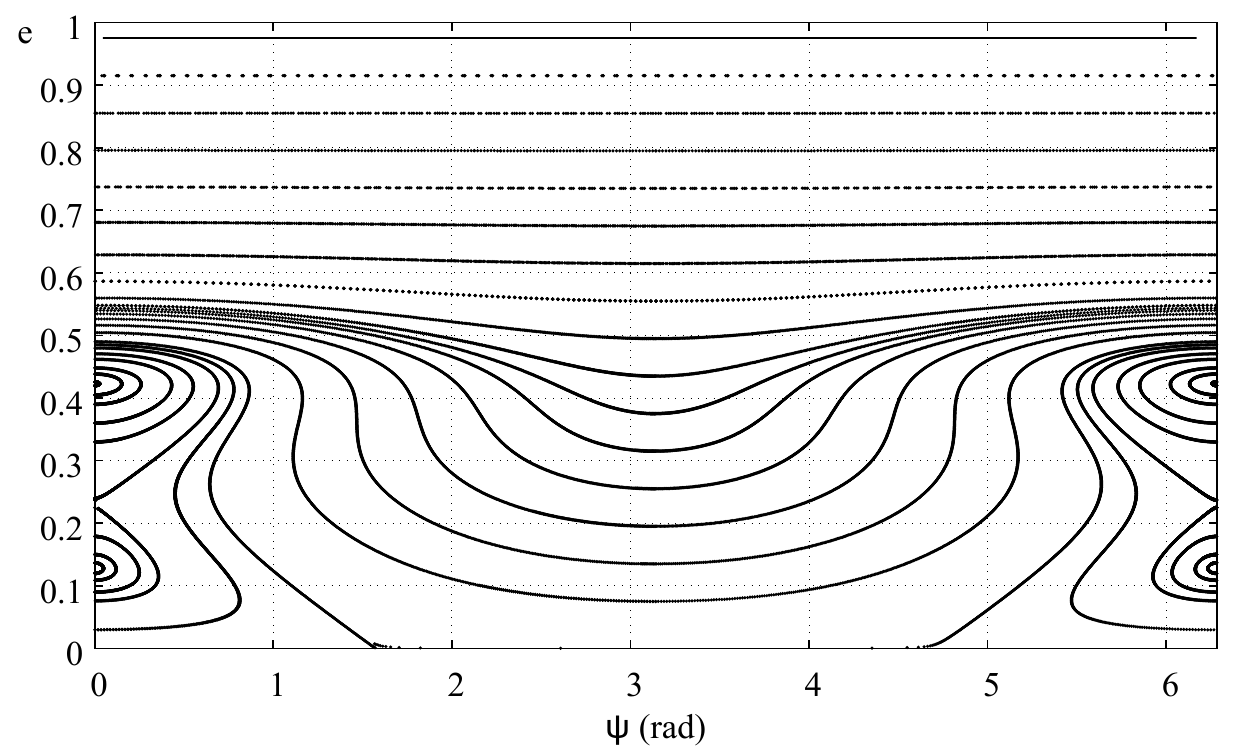} \\
  \includegraphics[width=0.49\textwidth]{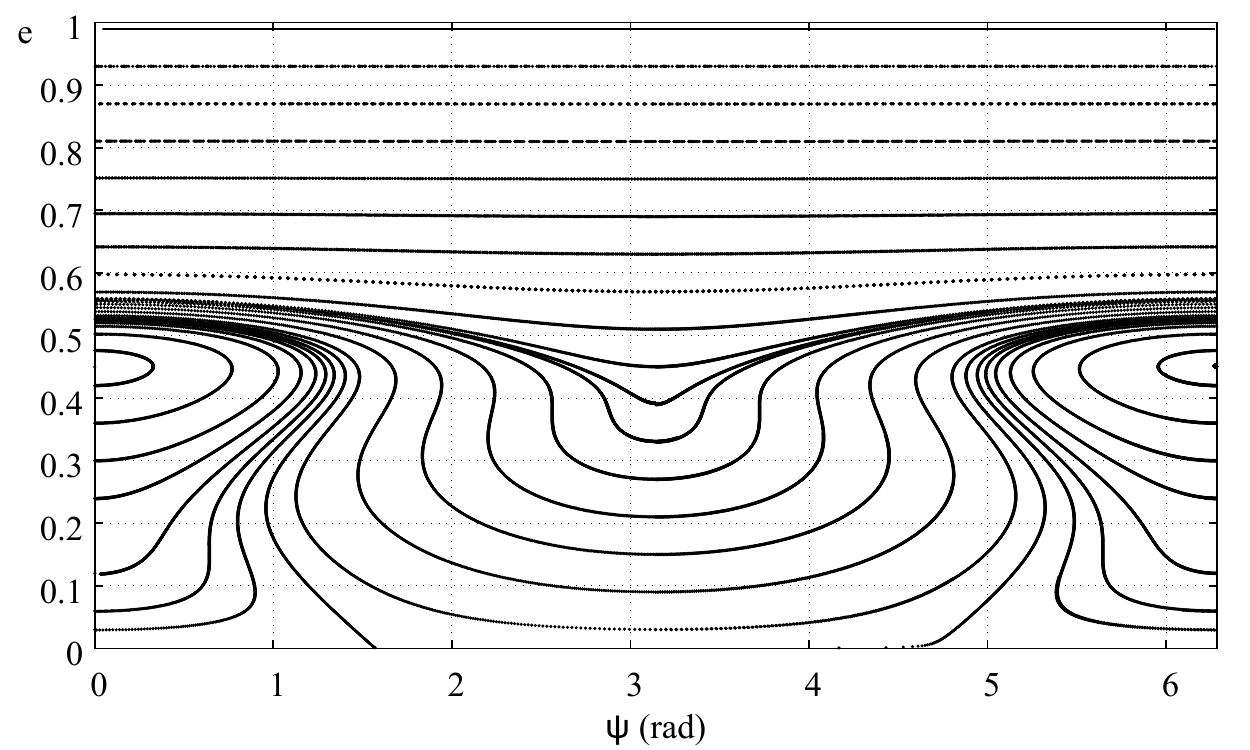}   \includegraphics[width=0.49\textwidth]{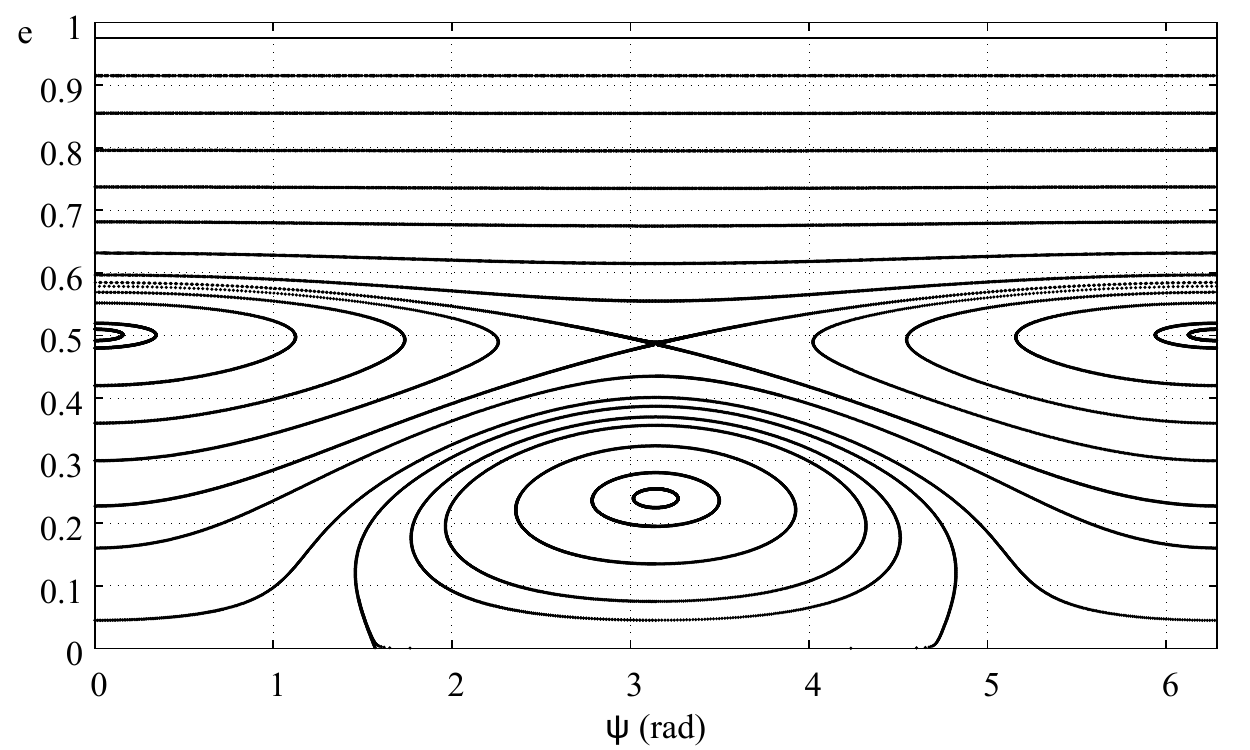} 
\caption{The possible phase space portraits for $j=1$, $A/m=1$ m$^2/$kg and $a=8078$ km, as indicated in Fig.~\ref{fig:esempio_Lambdae_res1_8078_a2m1} right. Top left: $\tilde{\Lambda}=-20.6$ km$^{1/2}$; top right: $\tilde{\Lambda}=-20.5$ km$^{1/2}$; bottom left: $\tilde{\Lambda}=-20.45$ km$^{1/2}$; bottom right: $\tilde{\Lambda}=-20.3$ km$^{1/2}$.}
\label{fig:quattrocasi_res1_8078_a2m1}       
\end{figure*}

The first example is presented in Fig.~\ref{fig:esempio_Lambdae_res1_8078_a2m1}. For $j=1$, $A/m=1$ m$^2/$kg and $a=8078$ km, the value of $\tilde{\Lambda}\equiv\Lambda/\sqrt{\mu}$ computed at the equilibrium points $\psi_1=0$ and $\psi_1=\pi$ is depicted as a function of the corresponding value of eccentricity, considering the solution for $\cos{i}$ on the first column of Fig.~\ref{fig:aei_res1}. We display $\tilde{\Lambda}$ instead of $\Lambda$ for the sake of clarity to avoid large numbers. In blue, the stable equilibrium points at $\psi_1=0$, in red the unstable ones. In green the stable equilibrium points at $\psi_1=\pi$, in ochre the unstable ones. On the right, a closer view is given, indicating also the number of equilibrium points existing in the different regions. In Fig.~\ref{fig:esempio_Lambdaei_res1_8078_a2m1}, the same information is showed, but displaying also the corresponding value of inclination (dashed lines, secondary $x$-axis). Following the two figures, we can see that for $j=1$, $A/m=1$ m$^2/$kg and $a=8078$ km, the phase space can be structured according to a minimum of 1 equilibrium point and to a maximum of 3 equilibrium points, which can exist in the range $e\in [0:1]$, $i\in [4:64]$ deg. Note the narrow range of inclination, $i\in [39.8:40.8]$ deg, within which the equilibrium point at $\psi_1=0$ can be unstable. Moreover, up to $\tilde{\Lambda}\simeq-20.55$ km$^{1/2}$ there exists only one equilibrium point at $\psi_1=0$, which is stable (see Fig.~\ref{fig:quattrocasi_res1_8078_a2m1} top left); from $\tilde{\Lambda}\simeq-20.55$ km$^{1/2}$ to $\tilde{\Lambda}\simeq-20.48$  km$^{1/2}$ there are 2 stable equilibrium points at  $\psi_1=0$ and one unstable equilibrium point at $\psi_1=0$  (see Fig.~\ref{fig:quattrocasi_res1_8078_a2m1} top right, note also the homoclinic connection); from $\tilde{\Lambda}\simeq-20.48$  km$^{1/2}$ to $\tilde{\Lambda}\simeq-20.44$ km$^{1/2}$ there exists one stable equilibrium point at $\psi_1=0$ (see Fig.~\ref{fig:quattrocasi_res1_8078_a2m1} bottom left); for values of $\tilde{\Lambda}$ higher than $\tilde{\Lambda}\simeq-20.44$ km$^{1/2}$, there exist one stable equilibrium point at $\psi_1=0$ and one stable and one unstable equilibrium points at $\psi_1=\pi$ (see Fig.~\ref{fig:quattrocasi_res1_8078_a2m1} bottom right). At the transitions, namely, at $\tilde{\Lambda}\simeq-20.55$ km$^{1/2}$, $\tilde{\Lambda}\simeq-20.48$ km$^{1/2}$ and $\tilde{\Lambda}\simeq-20.44$ km$^{1/2}$ there is a bifurcation: on the horizontal lines, for increasing  $\tilde{\Lambda}$, either there start appearing 2 additional equilibrium points, one stable and one unstable, or 2 existing equilibrium points with a different linear stability occur to coincide and then disappear.

\begin{figure*}
\centering
 \includegraphics[width=0.49\textwidth]{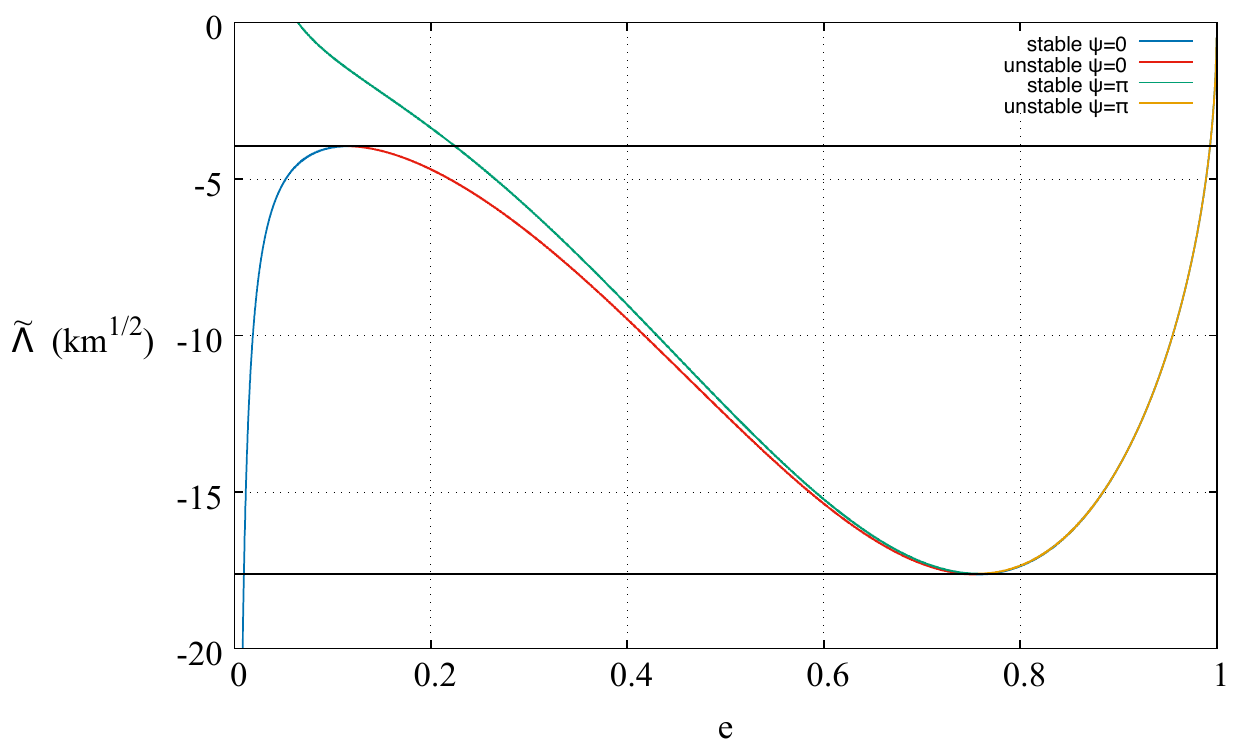}  \includegraphics[width=0.49\textwidth]{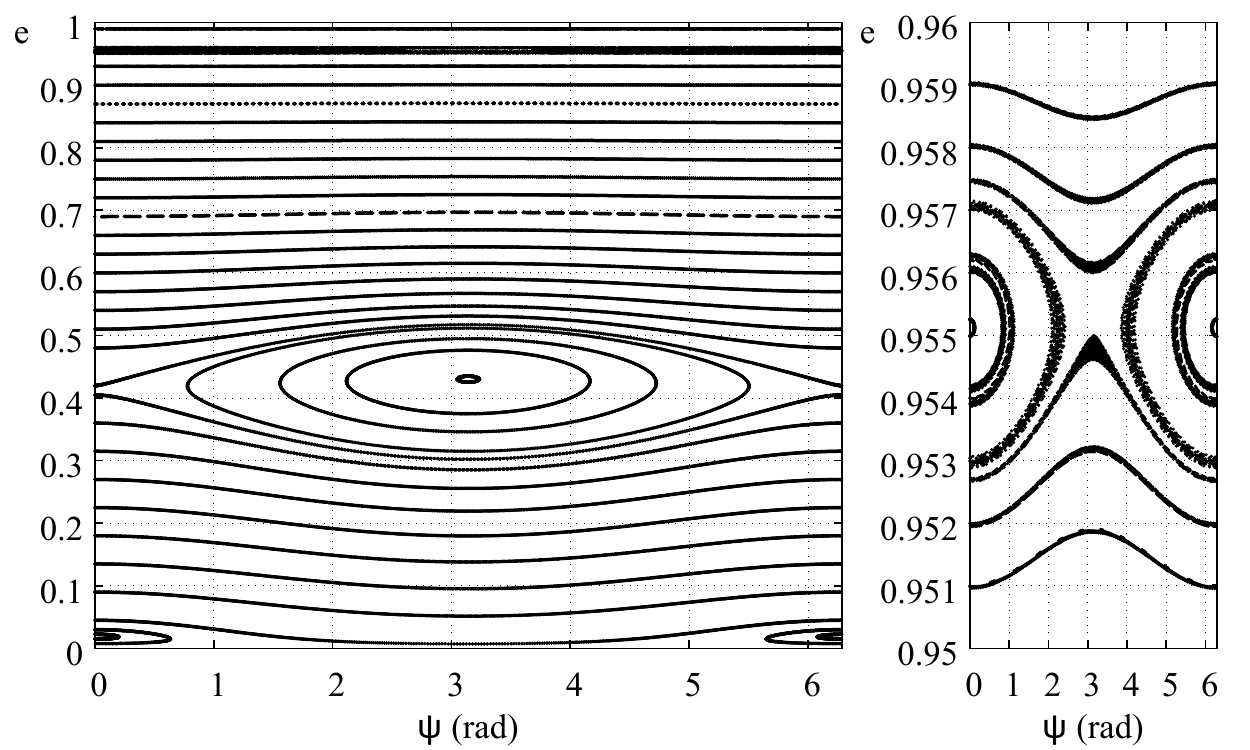} \\
\caption{On the left a close-up view of the $\tilde{\Lambda}$ behavior as a function of eccentricity, at the equilibrium points existing for $j=1$, $A/m=1$ m$^2/$kg and $a=12078$ km. Blue: stable point at $\psi_1=0$; red: unstable point at $\psi_1=0$. Green: stable point at $\psi_1=\pi$; ochre: unstable point at $\psi_1=\pi$. On the right, the phase space portrait structured in 5 equilibrium points for the same values of $a$, assuming $\tilde{\Lambda}=-10$ km$^{1/2}$. The right plot is split in two to highlight the phase space at $e\approx 0.955$.} 
\label{fig:esempio_Lambdae_res1_a2m1_5eq}       
\end{figure*}

Notice that Fig.~\ref{fig:esempio_Lambdaei_res1_8078_a2m1} can give also the information on the range of inclination within which we can have a given number of equilibrium points with the corresponding stability, and how the inclination changes according to the eccentricity, for given semi-major axis and $\tilde{\Lambda}$. In other words, the inclination corresponding to a given phase space portrait, such as the ones in Fig.~\ref{fig:quattrocasi_res1_8078_a2m1}, can be inferred by looking to Fig.~\ref{fig:esempio_Lambdaei_res1_8078_a2m1}. A preliminary definition of this domain was considered in \cite{SchettinoIAC}, taking into account only the strictly resonant behaviors for almost zero eccentricities. 

For $j=1$, $A/m=1$ m$^2/$kg, by increasing the semi-major axis, the number of possible equilibrium points can increase. Spanning a range of $a$ up to the GEO ring, the maximum number is 5, and it is found already at  $a=8178$ km. In Fig.~\ref{fig:esempio_Lambdae_res1_a2m1_5eq}, we show the corresponding behavior of the $(e,\tilde{\Lambda})$ curves and the phase space portrait associated with 5 equilibrium points for $a=12078$ km. Moreover, we note that  $(e,\tilde{\Lambda})$ curves can be not continuous, because Eq.~(\ref{eq:cosi}) sets specific constraints on the range of $\tilde{\Lambda}$ ($\cos{i}\in[-1,1]$).

Looking to the  $(e,\tilde{\Lambda})$ curves for $j=1$, always considering the solution for $\cos{i}$ on the first column of Fig.~\ref{fig:aei_res1}, for $A/m=20$ m$^2/$kg it is noticed a similar behavior, while for $A/m=0.012$ m$^2/$kg the number of equilibrium points can be 4 also in non-degenerate configurations. Taking as reference the phase space portrait depicted in Fig.~\ref{fig:esempio_Lambdae_res1_a2m1_5eq} on the right, for $A/m=0.012$ m$^2/$kg  and given $\tilde{\Lambda}$, the equilibrium point existing at the lowest eccentricity can disappear, and there persist one unstable equilibrium at $\psi_1=0$ and one stable equilibrium at $\psi_1=\pi$ corresponding to a same eccentricity, and one stable equilibrium at $\psi_1=0$ and one unstable equilibrium at $\psi_1=\pi$ corresponding to a different eccentricity.

With respect to the other resonant behaviors, it is stressed the following. Generally speaking, it appears that the dynamics associated with the resonant terms \#1, \#2 and \#4 is richer than for the other three resonant terms, in terms of maximum possible number of equilibrium points. Also, the value of area-to-mass ratio can change the value of semi-major axis at which a bifurcation can occur.  

\noindent\underline{Resonant term \#2.} For $A/m=1$ m$^2/$kg, considering the solution for $\cos{i}$ on the second column of Fig.~\ref{fig:aei_res2}, the curve corresponding to $\psi_2=0$ is higher than the curve corresponding to $\psi_2=\pi$, contrary to what happens for $j=1$. For the lower values of semi-major axis the maximum possible number of equilibrium points is 3: one stable at $\psi_2=\pi$ and one stable and one unstable at $\psi_2=0$, that is, as in Fig.~\ref{fig:quattrocasi_res1_8078_a2m1} on the bottom right, but the curves are shifted by $\pi$ in $\psi$.  At $a=9878$ km, it can appear also one unstable equilibrium at $\psi_2=\pi$ and the 3 equilibrium points can be as before or all associated to $\psi_2=\pi$, as in Fig.~\ref{fig:quattrocasi_res1_8078_a2m1} on the top right, but again shifted by $\pi$ in $\psi$. From about $a=10078$ km the phase space can be structured around 5 equilibrium points, as in  Fig.~\ref{fig:esempio_Lambdae_res1_a2m1_5eq} on the right, but also shifted by $\pi$ in $\psi$. For $A/m=20$ m$^2/$kg, the same behavior is found, while for $A/m=0.012$  m$^2/$kg, the only difference is that, analogously to what happens for the resonant term \#1, it is possible to have 4 equilibrium points, as the one associated with the lowest eccentricity drifts as much towards $e=0$ as to disappear. 

 \noindent\underline{Resonant term \#5.} The dynamics is analogous to what described for $j=1$ except that we do not ever notice 5 equilibrium points, and that there exist large regions where the number is only 2: one stable at $\psi_5=0$ and one unstable at $\psi_5=\pi$, or vice versa. In particular, for $A/m=0.012$  m$^2/$kg and $A/m=1$  m$^2/$kg, this is always the case from about $a=19000$ km.
 
\noindent\underline{Resonant term \#6.} The dynamics is analogous to what described for $j=2$, but, analogously to the case $j=5$, the maximum number of equilibria is 3 and in many cases we can see only 2.  In particular, for the three values of area-to-mass explored this is always the case from about $a=16500$ km. 

\noindent\underline{Resonant term \#3.} For $A/m=1$ m$^2/$kg, there are 3 equilibrium points: one stable equilibrium point at $\psi_3=0$ at a low eccentricity, and one unstable at $\psi_3=0$ and one stable at $\psi_3=\pi$. This occurs until about $a=15500$ km (about $a=15000$ km for  $A/m=0.012$ m$^2/$kg and about about $a=17000$ km for  $A/m=20$ m$^2/$kg), where the equilibrium at the lowest eccentricity disappears. From that value of semi-major axis on, the 2 equilibrium points drift towards higher values of eccentricity, analogously to what happens for the other resonant terms.

\noindent\underline{Resonant term \#4.} In this case, the dynamics can be as rich as in the case of resonant terms \#1 and \#2, in the sense that there can exist up to 5 equilibrium points: one stable at $\psi_4=0$ at a low eccentricity, one unstable at $\psi_4=0$  and one stable at $\psi_4=\pi$, and then one stable at $\psi_4=0$  and one unstable at $\psi_4=\pi$ at increasing eccentricity. This occurs for $A/m=0.012$ m$^2/$kg until about $a=10000$ km, for $A/m=1$ m$^2/$kg until about $a=11600$ km, for $A/m=20$ m$^2/$kg until about $a=19000$ km. From that value on, there persist 4 equilibrium points, which are associated to increasing values of eccentricity for increasing values of semi-major axis.

\section{Discussion and Future Directions}
\label{sec:disc}

The analysis proposed in this paper provides the fundamental ingredients to describe the dynamics of a body with a high area-to-mass ratio, which orbits a given major oblate body,  and which is subject to solar radiation pressure effects. 

\begin{figure*}
\centering
 \includegraphics[width=0.49\textwidth]{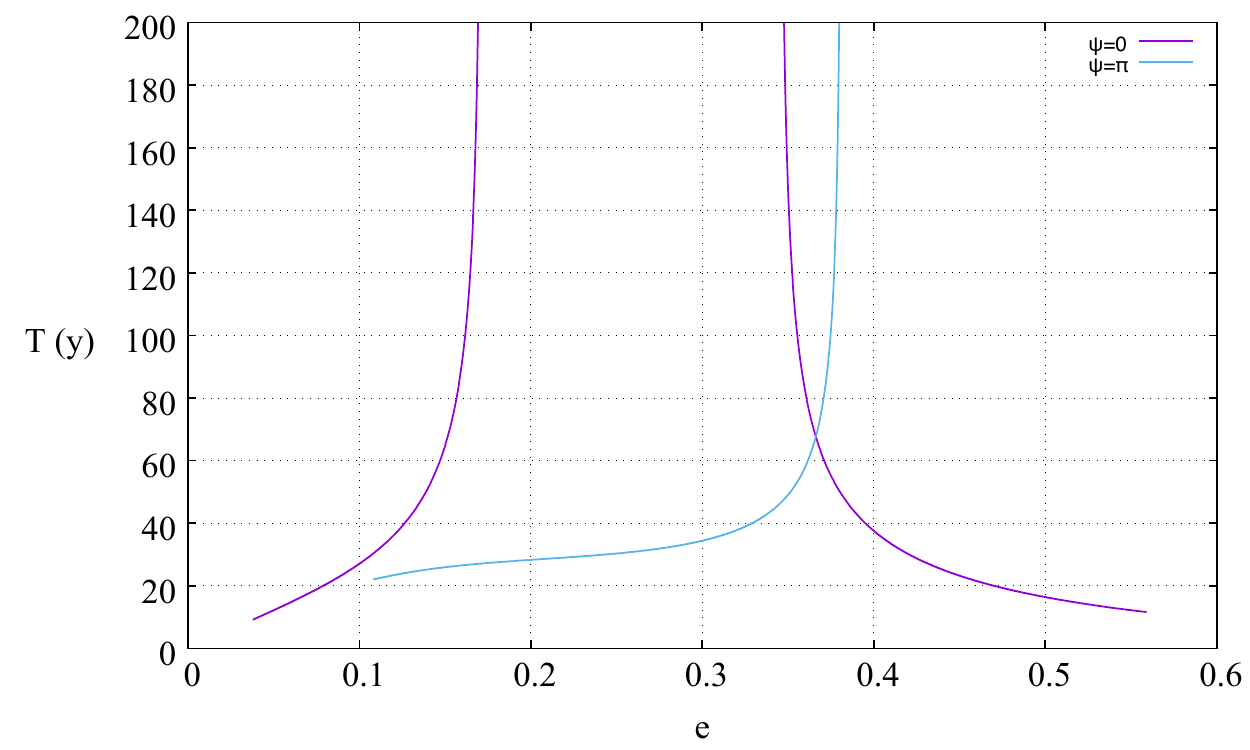}   \includegraphics[width=0.49\textwidth]{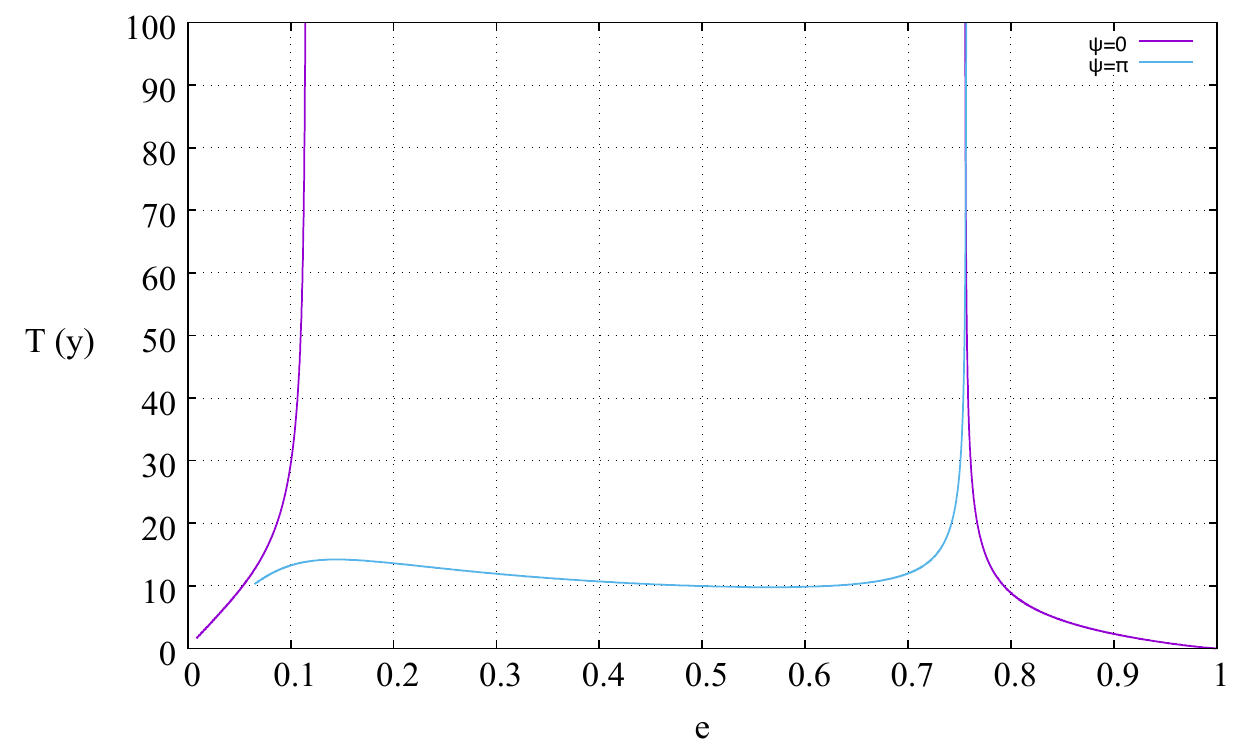}  
\caption{For the elliptic points existing for the resonant term \#1 and  $A/m=1$ m$^2/$kg, we show the period $T$ in years corresponding to the libration curves in their neighborhood. Left: $a=8078$ km, $\tilde{\Lambda}\in[-21:-20]$  km$^{1/2}$; right: $a=12078$ km, $\tilde{\Lambda}\in[-20:10]$  km$^{1/2}$.}
\label{fig:es_period_res1_a2m1}       
\end{figure*}

In particular, it is possible to apply all the dynamical systems theory tools (e.g., \cite{PC,W}) which can characterize the elliptic (stable) and hyperbolic (unstable) behaviors associated with the equilibrium points of the system. The information given by the eigenvalues and eigenvectors of the Jacobian matrix,  Eq.~(\ref{eq:jac}), computed at the corresponding equilibrium point, can be used to compute the invariant librating curves in the neighborhood of the elliptic points and the hyperbolic invariant manifolds in the neighborhood of the hyperbolic points. Concerning the time required to move along a curve, at an elliptic point the two eigenvalues are conjugate pure imaginary, say $\pm i\nu$. The closer the libration curve to the elliptic point, the closer its period to $T=2\pi/\nu$. In Fig.~\ref{fig:es_period_res1_a2m1}, we show the value of the period $T$ in years for the cases showed in Figs.~\ref{fig:esempio_Lambdaei_res1_8078_a2m1}-\ref{fig:esempio_Lambdae_res1_a2m1_5eq}. The asymptotic behavior that can be seen in the figure corresponds to the line of transition between different behaviors - the point of bifurcation described before. As the distance with respect to the elliptic point increases, the time needed to cover the curve also increases, in the limit to reach the hyperbolic manifold, which tends, by definition, asymptotically to the unstable point.  Note that the resonant curves corresponding to $\psi_j=\pi/2$ and $\psi_j=3\pi/2$, Eqs.~(\ref{eq:res})-(\ref{eq:coef_res}), always act as separatrices between libration and circulation motion, when an equilibrium point at low eccentricity exists.

The initial phase with respect to the Sun, that is, the longitude of the ascending node, the argument of pericenter and the epoch, is crucial, given initial $(a,e,i,A/m)$ to understand the dynamics that the small body can follow. In the neighborhood of one or more equilibrium points, it could be possible to play with the initial phase to move along a given invariant curve in the libration region or along a given separatrix. According to this, the eccentricity evolution will change and so the corresponding time will. Notice (see, e.g., Fig.~\ref{fig:quattrocasi_res1_8078_a2m1}) that even in case of circulation the natural dynamics can provide an eccentricity increase that can be exploited conveniently. With the same idea, it can be envisaged to play with the constants which determine the dynamics, $a$, $A/m$ or $\Lambda$ (or $\tilde{\Lambda}$), to achieve a given objective, in particular to obtain a bounded eccentricity variation within a given timespan or an escape trajectory. 

Moreover, from the analysis provided in Sec.~\ref{sec:model}, the maximum eccentricity corresponding to a given invariant curve can be computed, as suggested in \cite{AC_IAC}, by simply looking for the following condition
 \begin{equation}\label{eq:red_2}
 \begin{aligned}
\dot e_{|j}&=0\\
{\frac{d\dot e_{|j}}{dt}}
&=n_2C_{SRP}\frac{\sqrt{1-e^2}}{na}\mathcal{T}_{j}\dot\psi_j\cos{\psi}_j<0,
 \end{aligned}
\end{equation}
which assumes that the maximum occurs at either $\psi_j=0$ or $\psi_j=\pi$.

On the other hand, each invariant curve, as the ones depicted in Figs.~\ref{fig:quattrocasi_res1_8078_a2m1}-\ref{fig:esempio_Lambdae_res1_a2m1_5eq}, is characterized by a constant Hamiltonian -- Eq.~(\ref{eq:ham}) -- and this information can be considered to compute a priori the maximum and minimum eccentricity that can be achieved. For instance, departing from the unstable direction associated with a saddle point, corresponding say to $\psi=0$, we can {\bf see if an eccentricity corresponding to $\psi=\pi$ can be attained and, in case, its value,} by inverting Eq.~(\ref{eq:ham}) as a function of $\Sigma$, or $\sqrt{1-e^2}$. A detailed analysis on how this can be done will be the focus of a future investigation.

\begin{figure*}
\centering
 \includegraphics[width=0.49\textwidth]{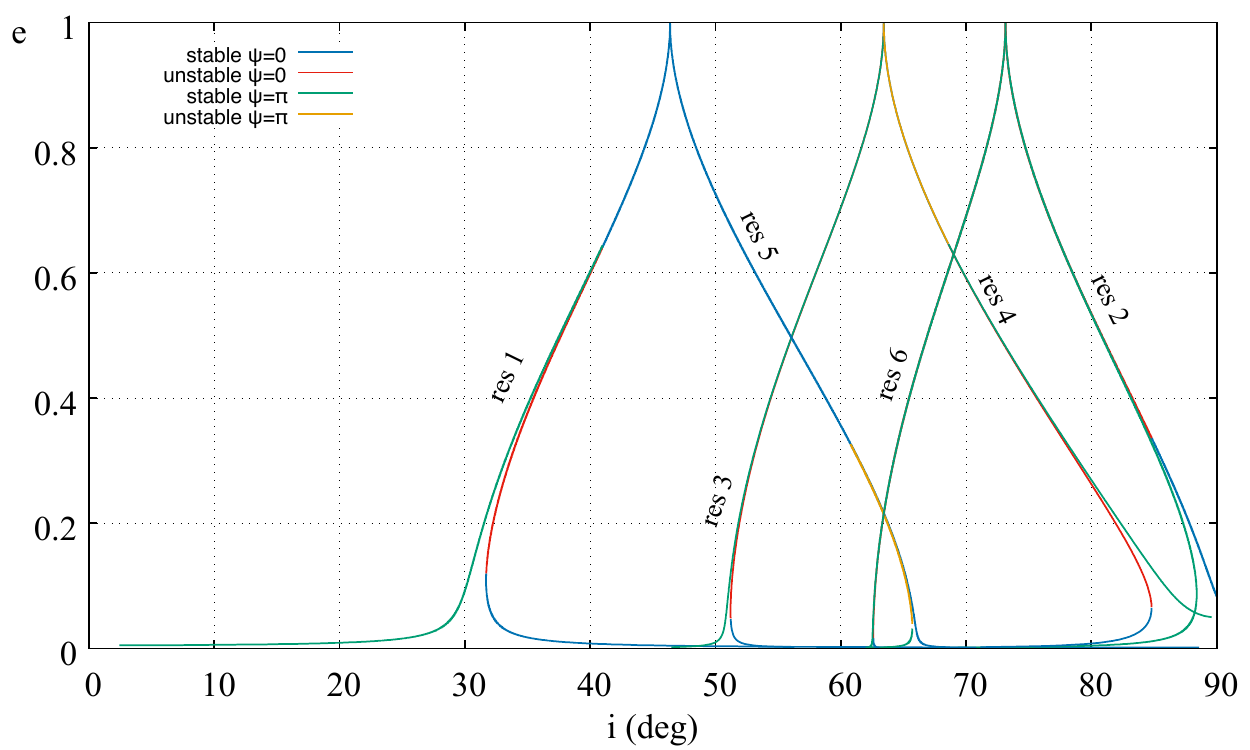}  \includegraphics[width=0.49\textwidth]{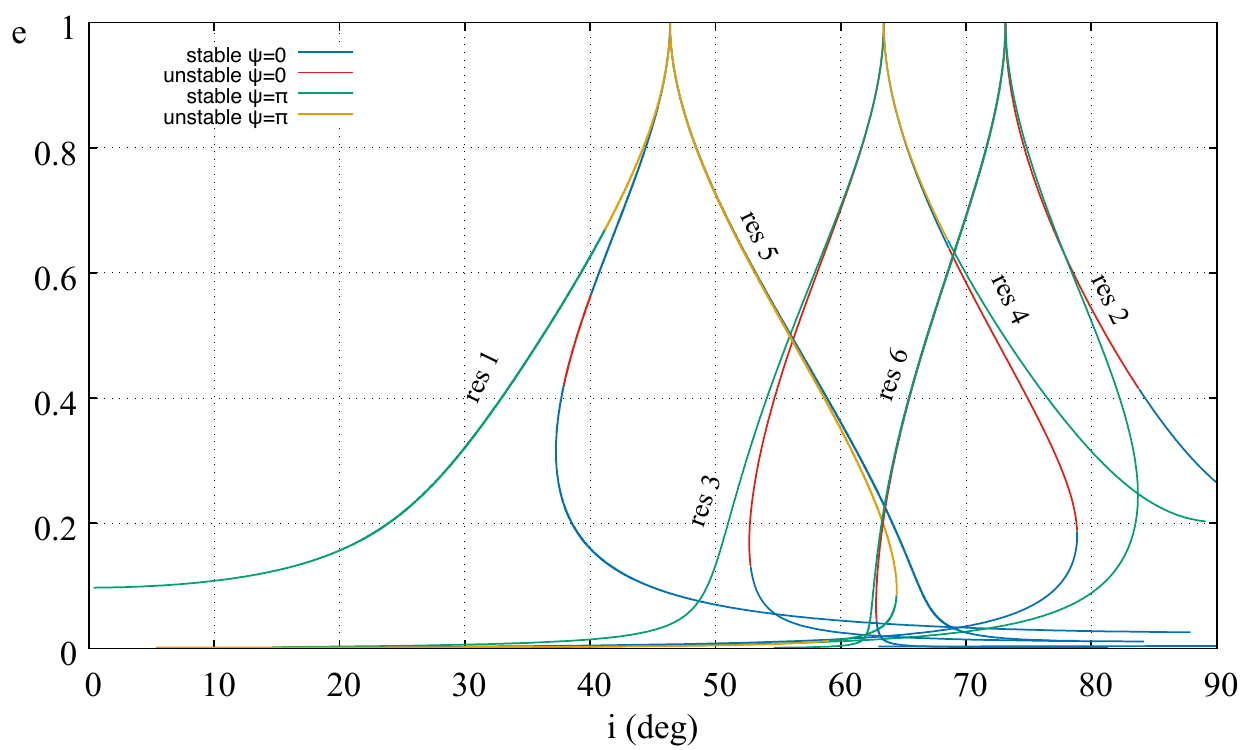}  
\caption{For $a=10078$ km, the curves showed represent the location of the equilibrium points, as a function of inclination and eccentricity for the six resonant terms. Blue: stable point at $\psi_j=0$; red: unstable point at $\psi_j=0$. Green: stable point at $\psi_j=\pi$; ochre: unstable point at $\psi_j=\pi$. Left: $A/m=1$ m$^2/$kg; right:  $A/m=20$ m$^2/$kg.}
\label{fig:es_inter}       
\end{figure*}

As it is well known, the single resonance hypothesis assumed in this work can fail to hold, that is, it can occur that two resonant terms can play a comparable role at the same time. This can be inferred from Figs.~\ref{fig:aei_res1}-\ref{fig:aei_res4}, which show that, for a same $(a,e,i,A/m)$ combination, there exist equilibrium points associated with different resonant terms. The same situation can be also depicted as an intersection, in the $(e,i)$ plane, between curves corresponding to equilibrium points for given $(a,A/m)$\footnote{as done for the resonant, in the strict sense, curves, e.g., in \cite{gC62,Alessi_MNRAS}.}. In Fig.~\ref{fig:es_inter}, we show an example for $a=10078$ km and $A/m=1$ m$^2/$kg and $A/m=20$ m$^2/$kg. Note how the stability can change by changing the area-to-mass ratio and also the number of intersections. By increasing the value of semi-major axis the curves tend to be parallel with respect to the $x-$axis towards higher values of eccentricity. 

\noindent{\bf The behavior in the regions where two dominant terms can exist will be studied in the future by analyzing the value of the corresponding amplitude and period and also the possible occurrence of a chaotic behavior.}

On the other hand, along an invariant curve of the phase space (recall Figs.~\ref{fig:quattrocasi_res1_8078_a2m1}-\ref{fig:esempio_Lambdae_res1_a2m1_5eq}) it is possible to estimate the excursion in eccentricity using Eqs.~(\ref{eq:red_2}) (and an analogous one to compute the minimum eccentricity achieved), and then apply Eq.~(\ref{eq:cosi}) to see the corresponding excursion in inclination \cite{Schettino_freq}. From such estimate, it would be trivial to see if a transition between different resonant terms might take place, by comparison with plots like those in Fig.~\ref{fig:es_inter}.

Future work will include the estimate of the size and location of the chaotic regions along with the corresponding Lyapunov time{\bf , not only in case of overlapping resonances but also in the neighborhood of the hyperbolic invariant manifolds associated with the hyperbolic equilibrium points}. Also, specific practical applications for both bounded and escape trajectories for the six resonant terms will be investigated. For instance, as shown in \cite{REF1}, the equilibrium points corresponding to orbits with apogee pointing the Sun can be exploited for enhanced Earth observation in the visible wavelength, while the equilibrium point corresponding to orbits with perigee pointing the Sun can be exploited for exploration mission of the Earth magnetosphere \cite{REF7,REF6}. The design of disposal trajectories with a solar sail, so far performed analytically only for planar orbit \cite{REF4} and numerically for inclined orbits \cite{REF15}, can be fully solved analytically with the method presented here. {\bf In the field of space debris, similarly to what was done in \cite{Volker} for the case of the ``resonant reentry corridors'', it will be of interest to analyze the space debris catalog to see whether there exists a specific case trapped in one of the resonances found. Other applications can be considered for different planetary systems, such as the frozen solutions found in \cite{Scheeres}.}

{\bf Finally, as mentioned at the beginning, it will be our priority to see how the phase space may change when an additional perturbation comes into play. In the case of the Earth, lunisolar gravitational perturbations and the ellipticity of the Earth will be considered for high-altitude orbits (see, e.g., \cite{CasanovaEtAl,GkoliasColombo,L1}) and the role of the atmospheric drag in the final deorbiting phase (see, e.g., \cite{R1}).}

\begin{acknowledgements}
The authors would like to acknowledge the funding received by the European Commission Horizon 2020, Framework Programme for Research and Innovation (2014-2020), under the
ReDSHIFT project (grant agreement n. 687500) and the funding received by the European Research Council (ERC) through the European Commission Horizon 2020, Framework Programme for Research and Innovation (2014-2020), under the project COMPASS (grant agreement n. 679086). The authors are grateful to Josep Masdemont, Ioannis Gkolias and Kleomenis Tsiganis for the useful discussions.
\end{acknowledgements}

\noindent

\noindent

{\it Compliance with ethical standards}

{\it Conflict of interest}: The authors declare that they have no conflict of interest.



\end{document}